\documentclass[preprint,12pt]{elsarticle}
\usepackage[english]{babel}
\usepackage[T1]{fontenc}
\usepackage{atbegshi}
\usepackage{natbib}
\usepackage{lineno,hyperref}
\usepackage{enumitem}
\usepackage{mathtools}
\usepackage[mathscr]{eucal}
\usepackage{dsfont}
\usepackage{fixltx2e}
\usepackage{array}
\usepackage{caption}
\usepackage{subcaption}
\usepackage{graphicx}%
\usepackage{hhline, multirow}%
\usepackage{amsmath}
\usepackage{amssymb}
\usepackage{amsfonts}
\usepackage{amsthm}
\usepackage{multicol}
\usepackage{mathrsfs}%
\usepackage[title]{appendix}%
\usepackage{enumitem}
\usepackage{adjustbox}
\usepackage{textcomp}%
\usepackage{manyfoot}%
\usepackage{booktabs}
\usepackage{tabularx}
\usepackage{array}
\usepackage{geometry}
\usepackage{algorithm}%
\usepackage{algorithmicx}%
\usepackage{algpseudocode}%
\usepackage{listings}%
\usepackage[hang,flushmargin]{footmisc} 
\usepackage{siunitx}
\usepackage{float}
\renewcommand{\arraystretch}{1.1} 

\usepackage[table]{xcolor}  

\usepackage{calrsfs}
\usepackage[
    left = \flqq{},%
    right = \frqq{},%
    leftsub = \flq{},%
    rightsub = \frq{} %
]{dirtytalk}
\newcolumntype{R}[1]{>{\raggedleft\arraybackslash}p{#1}}
\newcolumntype{L}[1]{>{\raggedright\arraybackslash}p{#1}}
\newcolumntype{P}[1]{>{\centering\arraybackslash}p{#1}}
\newtheorem{remark}{Remark}

\DeclareMathAlphabet{\pazccal}{OMS}{zplm}{m}{n}

\setcitestyle{square}

\journal{*}
\begin{document}
\begin{frontmatter}
\title{From \textit{fair} price to \textit{fair} volatility: \\Towards an Efficiency-Consistent Definition of Financial Risk}
\affiliation[first]{organization={Sapienza University of Rome},
            addressline={Via del Castro Laurenziano, 9}, 
            city={Rome},
            postcode={00161}, 
            country={Italy}}
            \cortext[cor1]{Corresponding author, sergio.bianchi@uniroma1.it}
\affiliation[fourth]{organization={University of Cassino and Southern Lazio},
            addressline={Via S. Angelo}, 
            city={Cassino},
            postcode={03043}, 
            country={Italy}}
\author[first]{Sergio Bianchi}\corref{cor1}
\author[first]{Daniele Angelini}
\author[first]{\\Massimiliano Frezza}
\author[fourth]{Augusto Pianese}

\begin{abstract}
Volatility, as a primary indicator of financial risk, forms the foundation of classical frameworks such as Markowitz’s Portfolio Theory and the Efficient Market Hypothesis (EMH). However, its conventional use rests on assumptions—most notably, the Markovian nature of price dynamics—that often fail to reflect key empirical characteristics of financial markets. Fractional stochastic volatility models expose these limitations by demonstrating that volatility alone is insufficient to capture the full structure of return dispersion. In this context, we propose pointwise regularity, measured via the Hurst–Hölder exponent, as a complementary metric of financial risk. This measure quantifies local deviations from martingale behavior, enabling a more nuanced assessment of market inefficiencies and the mechanisms by which equilibrium is restored. By accounting not only for the magnitude but also for the nature of randomness, this framework bridges the conceptual divide between efficient market theory and behavioral finance.
\end{abstract}



\end{frontmatter}



\section{Introduction and conceptual foundation}
\label{sec:Intro}
Volatility is a cornerstone in the realm of finance, wielding profound implications for investors, traders, and policymakers alike. Understanding and effectively managing volatility is crucial for risk assessment, portfolio diversification, and the overall stability of financial markets. It is also perceived as a vital indicator of market sentiment, often reflecting shifts in economic conditions, geopolitical events, and investor behavior. Yet, the widespread use of this concept often fuels the tendency to misuse the term <<volatility>> to mean <<riskiness>>, in the sense that the assets or markets with the highest volatility are inherently considered the riskiest. Interestingly, compared to academics, practitioners seem inclined to hold a clearer distinction in mind between the two notions\footnote{For example, in a renowned response during the 2007 Berkshire Hathaway annual meeting, Warren Buffet and Charlie Munger argued against the commonly-held belief that volatility is a measurement of risk. In Buffet's words: ``\textit{Volatility does not measure risk and the problem is that the people who have written and taught about volatility do not know how to measure risk and the nice thing about data which is a measure of volatility is that it is nice and mathematical and wrong in terms of measuring risk. It's a measure of volatility, but past volatility does not determine the risk of investing.}''}, and the underlying cause of this divergence runs deeper than it may appear. The contemporary origins of this perspective, which has become so pervasive as to seem almost axiomatic, can be discerned in the foundational work of \cite{Markowitz1952} of the Modern Portfolio Theory:

\begin{quote}
    \say{\textit{We next consider the rule that the investor does (or should) consider expected return a desirable thing and \textbf{variance of return an undesirable thing}. This rule has many sound points, both as a maxim for, and hypothesis about, investment behavior. We illustrate geometrically relations between beliefs and choice of portfolio according to the ``expected returns - variance of returns'' rule}}.
\end{quote}

Whatever measure is used in practice (see Table \ref{Volmeas}), assimilating volatility with risk is substantiated by certain modeling assumptions, which appear more controversial today than in past years or decades. They include the insensitivity to both the portfolio composition and the order in which data are generated.

\begin{table}[H]
\scriptsize
\centering
\caption{Comparison of Financial Volatility Measures}
\label{Volmeas}
\begin{tabular}{@{}llp{5cm}p{5cm}@{}}
\toprule
\textbf{Method} & \textbf{Type} & \textbf{Strengths} & \textbf{Weaknesses} \\
\midrule
Historical Volatility & Statistical & Simple and transparent; widely used benchmark & Backward-looking; does not incorporate forward-looking information or market sentiment \\
Implied Volatility & Market-based & Forward-looking; reflects market expectations embedded in option prices & Model-dependent (e.g., Black-Scholes); sensitive to moneyness and maturity (smile/surface effects) \\
Realized Volatility & Statistical & Empirically grounded; accurate when based on high-frequency data & Requires high-quality data; sensitive to market microstructure noise \\
GARCH Models & Model-based & Captures volatility clustering and persistence; accommodates conditional heteroskedasticity & Parameter-sensitive; may underreact to structural breaks or sudden shocks \\
Stochastic Volatility & Model-based & Realistic and flexible volatility dynamics; incorporates randomness in volatility itself & Computationally intensive; complex calibration; implies market incompleteness \\
VIX Index & Market-implied & Aggregates implied volatility across strikes and maturities; standardized market reference & Reflects expected, not realized volatility; may be influenced by option market inefficiencies \\
\bottomrule
\end{tabular}
\end{table}

Regarding the first point, volatility can be considered desirable rather than undesirable when portfolios include variance derivatives or derivative strategies, such as for example straddles or strangles, which become profitable as the underlying asset's volatility increases\footnote{Consider for example the case of a long straddle or a long strangle. These options strategies involve buying a call and a put having the same underlying asset, the same expiration date and the same strike price (straddle) or different strike prices (strangle). In both cases, the more the price of the underlying changes (regardless the direction) the more in-the-money the straddle (strangle) will be, and the holder can exercise the relevant option to earn a profit.}. In these cases, volatility should not be minimised, but instead maximised. This highlights that volatility, in and of itself, is not inherently advantageous or disadvantageous nor can measure the level of risk, when this is intended as the potential loss stemming from holding a specific position. Therefore, the notion of risk is not inherent in the \textit{variability} of a phenomenon, but rather in its \textit{unpredictability}. As long as the financial world was perceived as a simple dichotomy, with chance (efficient markets and martingales) and determinism (nonlinear dynamical systems) being the only two opposing alternatives, the question of predictability did not even arise, because market efficiency, with its Brownian martingales-based paradigmatic models, implied the completely erratic nature of data, and the consequent interchangeability between risk and volatility. Over the past four decades, particularly in the field of physics, there has been a growing recognition of the continuum of degrees of randomness that bridges chance and determinism. Although this awareness has also permeated finance, it has not yet led to the consequential reevaluation of the concept of risk, which remains largely anchored to the notion of volatility as if <<dispersion>> was the only aspect that matters. Actually, dispersion is only one part of the story, which should also account for regularity, that is the intensity of smoothness (roughness) of the observed dataset. Unlike dispersion measured by volatility, regularity is sensitive to the order in which data are generated. 

\begin{figure}[H]
    \centering
    \includegraphics[width=0.7\textwidth]{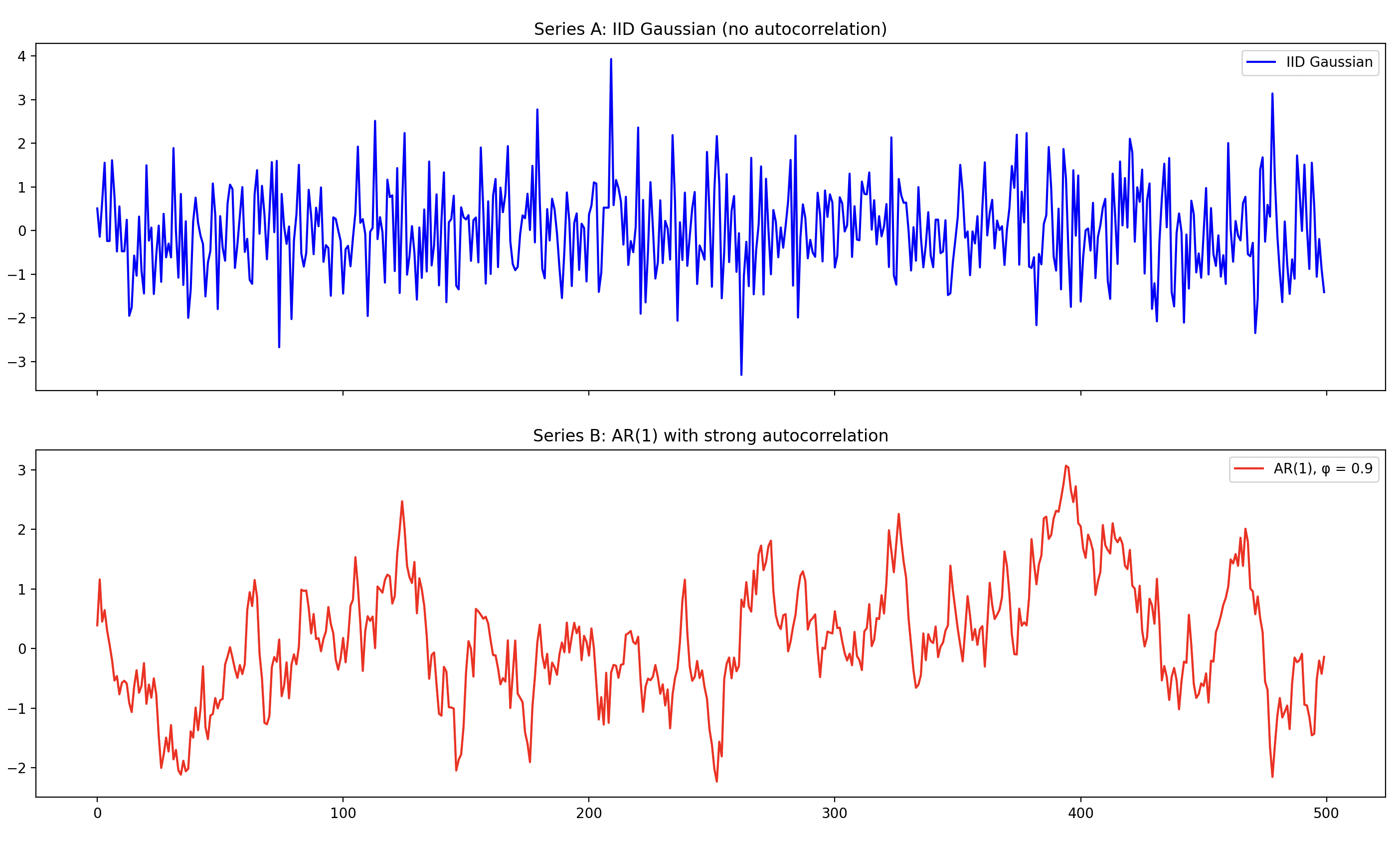}
    \captionsetup{margin=1cm}
    \caption{Comparison of two return series with identical volatility: IID Gaussian (top) vs. AR(1) with $\phi = 0.9$ (bottom)}
    \label{fig:RV1}
\end{figure}
This argument is exemplified in Figure \ref{fig:RV1}, where two synthetic return series with the same volatility (i.e., same standard deviation of increments), but drastically different dependence structures are displayed: series A (top panel) represents a white noise process. Returns are independent and identically distributed (IID), exhibiting no autocorrelation. There is no persistence or recognizable pattern over time; series B (bottom panel) is an AR(1) with a strong positive autocorrelation parameter $\phi = 0.9$. This causes significant persistence: large returns tend to follow large returns of the same sign.\\
Despite having the same volatility, the two sequences exhibit very different dynamics. Volatility measures the magnitude of dispersion, but not its temporal structure. It is blind to the presence or absence of autocorrelation, which significantly influences predictability and perceived risk (see, e.g., \cite{Sobolev2016} for the risk perceived by traders with respect to this kind of patterns). Since volatility and autocorrelation capture distinct aspects of a time series, a complete risk assessment requires tools that also account for the \textit{dependence structure}, not merely the second moment.

\begin{figure}[H] 
 \centering
\includegraphics[scale=.15, trim=0pt 0pt 0pt 0pt]{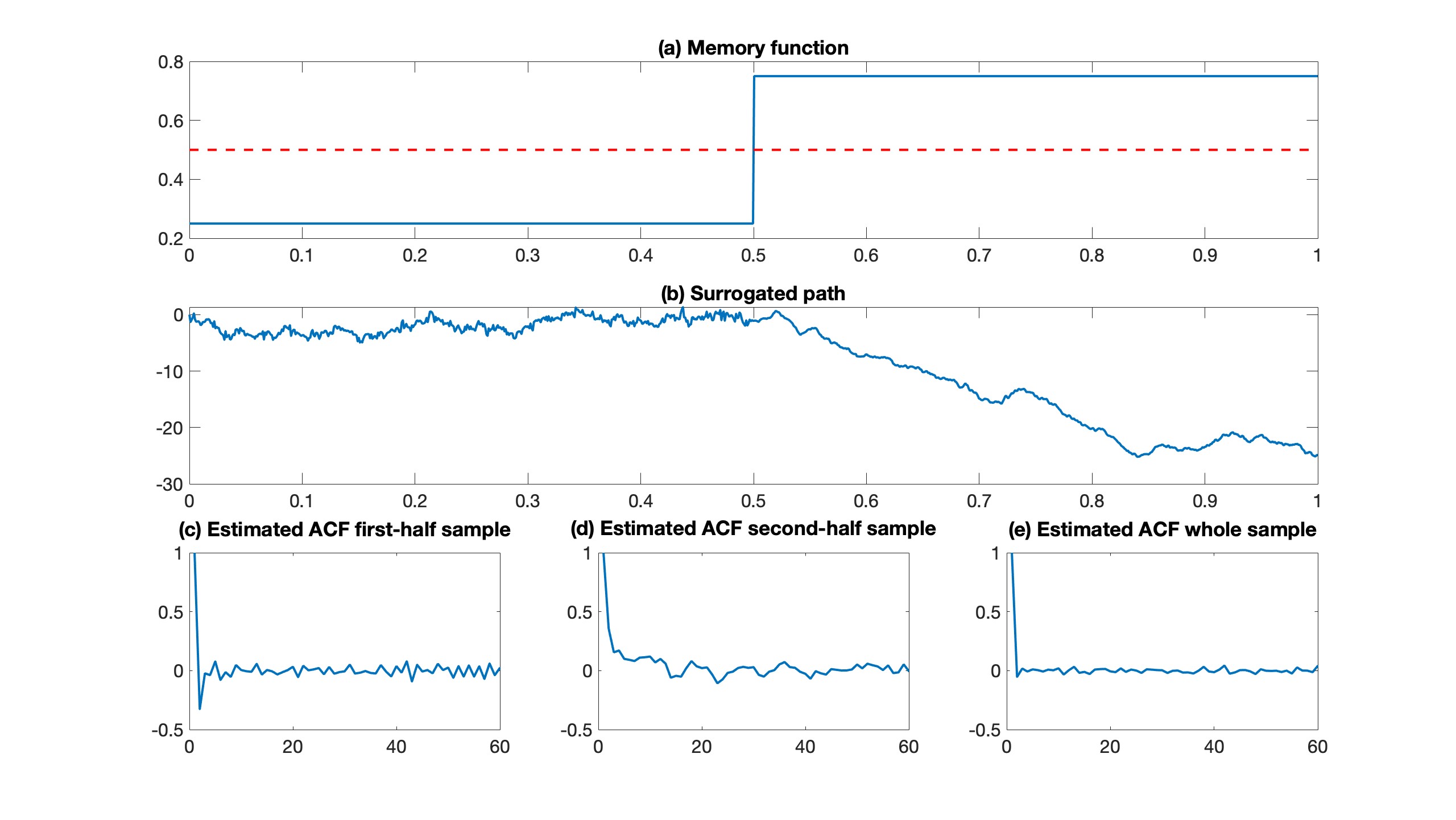}
\captionsetup{margin=1cm}
\caption{(a) step memory function (values larger than 0.5 indicate positive autocorrelation, values smaller than 0.5 negative autocorrelation, 0.5 indicates independence); (b) surrogated path of the stochastic process; bottom panels: autocorrelation of the increments of the first half of the path (c), of the first half of the second half of the path (d), and of the overall path (e).}
\label{fig:ACF1}
\end{figure}

This example, engineered by manipulating the autocorrelation function to fine-tune the smoothness of trajectories, elucidates the distinction between \textit{volatility} and \textit{regularity}: the former quantifies "\textit{how much}" dispersed data are around their mean value, while the latter captures "\textit{how}" data are dispersed. In the (Brownian) martingales framework, the "\textit{how}" is inconsequential, because independence univocally determines the behavior of the quadratic variation. However, when this paradigmatic model is challenged or relaxed, complexities emerge that necessitate disentangling the notions of volatility and regularity to prevent a wholesale conflation of volatility with risk. Alterations in the autocorrelation function clearly influence the regularity of the sequence and introduce potential errors in the evaluation of financial risk when this solely relies on volatility. The fact that empirical analyses extensively document negligible autocorrelation for financial return series is not in itself conclusive, nor does it diminish the relevance of the question, since a zero autocorrelation may appear as a consequence of positively and negatively autocorrelated data or, more generally, of nonstationarity. This effect - particularly significant for stochastic volatility models with memory - can be illustrated by the example in Figure \ref{fig:ACF1}, where (top panel) a step memory function is used to surrogate a trajectory (mid panel) of a stochastic process whose regularity is dictated by the values of the memory function. The increments of the first half of the trajectory exhibit a short-term negative autocorrelation (bottom left panel), whereas the increments of the second half exhibit a slowly decaying positive autocorrelation (bottom mid panel). When one calculates the autocorrelation for the entire sequence of increments (bottom right panel), this appears to be nearly zero, and this would misleadingly suggest that the data are uncorrelated. Clearly, the example represents an exaggerated illustration of the mechanism: reality is subtler than depicted, because the subsamples of values with varying degrees of autocorrelation can be shorter and differ in length.
The above considerations suggest that non-stationarity constitutes another major limitation to the use of volatility as a measure of financial risk.\\

Another limitation stems from volatility's inherently relative nature: it can only be meaningfully interpreted in comparison to its historical values, making it challenging to assess whether the current level of volatility is appropriate or not. In contrast to prices, the notion of a <<fair>> volatility — which could offer practical and predictable implications — does not exist and cannot be formally defined within the current theoretical framework.\\

Also coherent risk measures 
-- such as \textit{Value at Risk (VaR)} or \textit{Expected Shortfall (ES)} -- display limitations that can be partially traced back to what was observed earlier: 
\begin{enumerate}\setlength\itemsep{.25em}
\item (Dependency on Distributional Assumptions) Like volatility, coherent risk measures often assume a specific probability distribution of returns. If the distribution is misspecified, the risk measure becomes unreliable.
\item (Lack of Sensitivity to Temporal Structure) Just as volatility ignores the order of returns, so do most coherent risk measures, which typically summarize distributional outcomes without accounting for path-dependence or data regularity (e.g., two price paths with identical ES may exhibit vastly different short-term predictability or perceived risk).
\item (Aggregation and Subadditivity Assumptions Can Be Misleading) While \textit{subadditivity} is a desirable mathematical property, it can be problematic when agents or portfolios are not additive in behavior. This happens for example when one considers derivative strategies where risk is nonlinear in volatility. Therefore, risk aggregation can mask the true risk exposure of highly nonlinear portfolios.
\item (Lack of Absolute Benchmark or <<fair risk>>) also coherent risk measures do not provide a notion of fair risk or volatility. They quantify downside risk but cannot prescribe when the level of risk is <<adequate>> in a normative sense. This limits their use in strategic allocation and macroprudential regulation.
\item (Non-Uniqueness and Model Risk) For the same portfolio, different coherent risk measures (e.g., VaR vs. ES at different confidence levels) yield different rankings and decisions and the fact that no canonical coherent risk measure exists introduces subjectivity, especially in stressed or nonstationary regimes.
\item (Failure to Capture Liquidity Risk, Market Impact, or Feedback Loops) Coherent risk measures treat losses as exogenous, whereas real markets involve endogenous risk from feedback and reflexivity.
\end{enumerate}

The aforementioned circumstances necessitate a critical rethinking of the quantification of financial risk. Considering that this is related to the degree of unpredictability rather than the degree of dispersion, it is pertinent to investigate whether:
\begin{itemize}
  \item[(a)] measures exist that are capable to quantify the predictability, and
    \item[(b)] such measures can outdo the limitations of volatility discussed earlier.
\end{itemize}
Both questions are addressed in this work in the light of a new model which represents the most immediate and powerful generalization of the paradigmatic Brownian motion: the \textit{Multifractional Process with Random Exponents} (MPRE). This process acts on the autocorrelation structure in such a way that the pointwise regularity of its trajectories is governed by the functional parameter $H(t)$, referred to as the \textit{Hurst–Hölder exponent} at time $t$. As will be shown, through the behavior of $H(t)$, the MPRE can capture and reproduce the effects of a time-varying volatility, including as special cases those models which do not exhibit memory (e.g. Heston or SABR models). We will prove that an analytical relationship can be derived between the value of the exponent $H(t)$ and the realized volatility at the same time $t$. We will also show that the empirical validity of this relationship provides strong support for the MPRE as a realistic model of log-price dynamics. The relevance of the model extends beyond purely mathematical or statistical considerations: it is in fact possible to associate values of $H(t)$ with different phases of the market and, through this association, to express volatility in terms of the level that would be expected under market efficiency. In other words, the relationship between $H(t)$ and the realized volatility makes it possible to:
\begin{itemize}
    \item[(i)] decouple the measurement of financial risk from the simplistic dispersion (as in standard volatility measures)
    \item[(ii)] assess whether a given level of volatility is or not \textit{fair}, that is, if it is consistent with the level that would be observed in an efficient market. Since markets tend to restore the equilibrium, the eventual discrepancy between the estimated volatility and its \textit{fair} value would provide information about how quickly the market will restore equilibrium (i.e., efficiency) after occasional deviations from it. 
\end{itemize}

The remainder of the paper is structured as follows. Section \ref{sec:Model} introduces the theoretical framework and derives the analytical relation between volatility and regularity within the context of the MPRE. Section \ref{sec:RiskVol} discusses the financial interpretation of this volatility-regularity relationship. Section \ref{sec:Hestimation} briefly reviews key properties of established estimators for the Hurst–Hölder exponent, while Section \ref{sec:Application} presents an empirical analysis that provides strong evidence in support of the proposed relationship. Finally, Section \ref{sec:Conclusion} offers concluding remarks.

\section{Framework}
\label{sec:Model}
\subsection{Preliminaries}
Efficient Market Hypothesis (EMH) \citep{F1970} implies that stock prices fully reflect all available information. In its semi-strong form and given a filtered probability space \((\Omega, \pazccal{F}, (\pazccal{F}_t)_{t \in \mathbb{R}}, \mathbb{P})\), this aligns with the idea that the discounted price process $\hat{S}_t$ should be a martingale under the real-world probability measure $\mathbb{P}$ (or more commonly, under a risk-neutral measure $\mathbb{Q}$ when accounting for discounting), i.e. 
\begin{equation*}
    \mathbb{E}^{\mathbb{Q}}(\hat{S}(t)|\pazccal{F}_{\tau})=\hat{S}(\tau) \quad \quad (\tau \leq t).
\end{equation*}
The behavior of $\hat{S}(t)$ provides key insights into whether the process is or not a martingale: a) its total variation on any interval $[0,t]$ must be unbounded (except for trivial cases like constant processes); b) its quadratic variation $[\hat{S}]_t$ must be finite (for semimartingales, the martingale component is uniquely tied to the quadratic variation); c) the $p$-th variations, for $p > 2$, are typically zero (non-zero higher-order variations suggest deviations from martingale structure as, for example, for Hermite processes). Thus, while total variation is too coarse, and higher-order variations are typically degenerate, quadratic variation is the definitive tool for classifying martingales since it captures the cumulative unpredictability of increments. This underpins stochastic calculus (Itô's lemma) and no-arbitrage theory in finance.\\
When one looks at the Brownian motion paradigm, things become more specific. Denoted by \(\{B(t)\}_{t \in \mathbb{R}}\) the standard Brownian motion and by $\sigma$ a constant volatility, $X(t) = \sigma B(t)$ is a Brownian martingale. 
As a consequence of the L\'evy's Characterization Theorem \cite{Levy1939}, denoted by $X(t^-)$ the left limit of $X(t)$ with respect to $t$, the quadratic variation of $X(t)$ over an interval $[0, t]$ is of course $[X](t) =\sum_{0<s\leq t}(X(s)-X(s^-))^2= \sigma^2 t.$ Thus, the realized volatility $\sigma$ can be expressed in terms of the quadratic variation $[X](t)$ as \cite{Barndorff2002}
\begin{equation}\label{eq:VolQuad}
    \sigma = \sqrt{\frac{[X](t)}{t}}.
\end{equation}
Thus, since the (ir)regularity of Brownian motion is entirely captured by its quadratic variation, the most natural candidate for characterizing risk is volatility. In this way, within the paradigm of efficient markets, the consideration of risk measures beyond volatility becomes unnecessary.\\

Despite this rather simplified framework, it is a well-established fact that volatility is not constant over time, and therefore Brownian motion does not provide a realistic model for the dynamics of logarithmic asset prices in financial markets. As a consequence, literature has in depth explored stochastic volatility models (e.g., Heston model, SABR model, or a general Itô process with stochastic diffusion term) which can indeed be consistent with the martingale property, provided that some conditions (Novikov’s condition, Kazamaki’s and Feller's condition for CIR processes) are met. In particular, a) the drift of the discounted price process must be zero (or appropriately compensated in the risk-neutral measure) and b) the volatility process itself must be adapted and should not introduce arbitrage. Thus for a general Itô process $X_t$ with stochastic volatility $\sigma_t$ such that $dX(t) = \sigma(t) dB(t)$, the quadratic variation is an integral of the squared volatility process (or integrated variance), provided that it exists finite
\begin{equation*}
    [X](t)=\int_0^t \sigma^2(\tau) d\tau <\infty.
\end{equation*}
Of course, if $\sigma(t)$ is stochastic, the quadratic variation of $\hat{S}(t)$ becomes random, i.e.
\begin{equation*}
    \mathbb{E}^{\mathbb{Q}}\left([\hat{S}](t)\right)=\mathbb{E}^{\mathbb{Q}}\left(\int_0^t \sigma^2(\tau)\hat{S}^2(\tau) d\tau\right),
\end{equation*}
and if $\sigma(t)$ is mean-reverting (as for example in the Heston model), the long-term average of the quadratic variation may stabilize, but it remains path-dependent.\\
Since realized variance is an estimator of $[\hat{S}](t)$, the difference between realized variance and expected quadratic variation drives volatility trading strategies (e.g., variance swaps, VIX products).\\
Whether using simple Brownian martingales or more sophisticated stochastic volatility models driven by (eventually correlated) Brownian motions, randomness in returns is solely characterized by the growth of the quadratic variation, proportional to the length of the time interval $t$ in the former or given by the integrated variance in the latter. This exclusive relationship provides a theoretical foundation for using volatility as a measure of risk, and for scaling it over time according to equation (\ref{eq:VolQuad}). A well-known implication of this principle is the common practice of annualizing daily volatility by multiplying it by the square root of 252—the approximate number of trading days in a year. The same square-root-of-time rule underlies regulatory frameworks such as Basel II, Basel III, and guidelines issued by the U.S. Federal Reserve, which prescribe scaling the one-day Value at Risk (VaR) by the square root of ten to estimate the VaR over a ten-day (or two-week) holding period.\\

An advancement beyond this framework involves considering processes in which the stochastic volatility exhibits memory, reflecting more accurately the behavior observed in real financial markets. Such processes, which take the general form
\begin{equation}\label{eq:stochvolfbm}
\begin{cases}
      dS(t)= \mu \, S(t)\, dt + \sigma(t) \, S(t) \, dB(t)\\
      d\sigma(t) = \alpha(\sigma(t),t)\,dt+\nu(\sigma(t),t)\,dB^H(t)\\ 
    \end{cases}
\end{equation}
incorporate an instantaneous volatility process $\sigma(t)$ whose dynamics is leaded by a \textit{fractional Brownian motion} (fBm) $B^H(t)$ with Hurst parameter $H$. When $H>0.5$ the model displays long memory and smoothness \citep{ComteRenault1998}, whereas for $H<0.5$ the model displays short memory and roughness \citep{Gatheral2018}. For example, if $\alpha(\sigma(t),t)=-\lambda \, \sigma(t)$ and $\nu(\sigma(t),t)=\nu$, $\sigma(t)$ follows a fractional Ornstein-Uhlenbeck process, where $\lambda > 0$ is the mean-reversion rate and $\nu > 0$ is the volatility of volatility.\\

Given that fBm is not a semimartingale \citep{R1997}, the framework introduced in \eqref{eq:stochvolfbm} prompts the question of the extent to which such models depart from the equilibrium implied by the EMH, along with the implications previously discussed. In order to investigate this issue and to construct a financial risk indicator that offers deeper insights into market dynamics in the presence of stochastic volatility, we consider the \textit{Multifractional Processes with Random Exponents} (MPRE) framework \citep{AT2005,AEH2018,Lobodaetal2021}. To illustrate the key features of this class of processes, we first review some fundamental properties of the fBm which will be useful in the sequel.

\subsection{Fractional Brownian motion (fBm)}
Introduced by \cite{Kolmogorov1940} and disseminated by \cite{MV1968}, the now classic Fractional Brownian Motion (fBm) of Hurst parameter $H \in (0,1)$ is a centered self-similar Gaussian process defined by the integral
\begin{eqnarray} \label{eq:fbm}
    B^H(t) &=& \frac{1}{\Gamma\left(H+1/2\right)}\left\{\int_{-\infty}^0[(t-s)^{H-1/2}-(-s)^{H-1/2}]dB(s)+\int_{0}^t[(t-s)^{H-1/2}]dB(s)\right\}\nonumber \\
    &=& \frac{1}{\Gamma\left(H+1/2\right)}
    \int_{-\infty}^t \left[ (t-s)_+^{H-1/2} - (-s)_+^{H-1/2} \right] dB(s),
\end{eqnarray}
where $(\cdot)_+ = \max(\cdot,0)$. The process has stationary and normally distributed increments with mean 0 and variance given by
\begin{equation} \label{eq:scalvar}
    \mathbb{E}[B^H(t + T) - B^H(t)]^2=V_H T^{2H}.
\end{equation}
Since the primary focus is the scaling law that links the time increment $T$ to the variance through the Hurst parameter $H$, the quantity $V_H$ is often overlooked at the extent that many authors (see, e.g. \citep{Mishura2008}) define the fBm as
\begin{equation} \label{eq:fbm01}
    \bar{B}^H(t):=\frac{\left[2H\sin(\pi H) \Gamma(2H)\right]^{1/2}}{\Gamma(H+1/2)}\int_{-\infty}^t \left[ (t-s)_+^{H-1/2} - (-s)_+^{H-1/2} \right] dB(s),
\end{equation}
where the quantity $\left[2H\sin(\pi H) \Gamma(2H)\right]^{1/2}$ is introduced to enforce the factor $V_H$ to be equal to $1$.\\
However, since by \eqref{eq:scalvar} $V_H$ can in fact be interpreted as the variance of unit-lag increments, it is a quantity of particular relevance for our analysis; it is therefore worthwhile to examine it more closely. One has (see e.g. \citep{Reed1995})
\begin{equation}
    V_H = \frac{1}{\Gamma\left(H+\frac{1}{2}\right)^2}\left\{\frac{1}{2H}+\int_1^{\infty}\left[u^{H-1/2}-(u-1)^{H-1/2}\right]^2\,du\right\}
\end{equation}
and since it can be proved that
\begin{equation*}
    \int_1^{\infty}\left[u^{H-1/2}-(u-1)^{H-1/2}\right]^2\,du=-\frac{1}{2H}-\frac{2\Gamma(H+\frac{1}{2})\Gamma(-2H)}{\Gamma(-H+\frac{1}{2})}
\end{equation*}
it follows that
\begin{eqnarray} \label{eq:VH}
    V_H &=& \frac{1}{\Gamma(H+\frac{1}{2})^2}\left[\frac{1}{2H}-\frac{1}{2H}-\frac{2\Gamma(H+\frac{1}{2})\Gamma(-2H)}{\Gamma(-H+\frac{1}{2})}\right] \nonumber \\
    &=& \frac{2}{\pi}\Gamma(-2H)\sin\left(H\pi-\frac{\pi}{2}\right) \quad \text{(reflection identity)}\nonumber \\
    &=& -\frac{2}{\pi}\Gamma(-2H)\cos(\pi H) \nonumber \\
    &=& \frac{\Gamma(1-2H)}{\pi H} \cos (\pi H).
\end{eqnarray}

\begin{remark} \label{rem:Remark1}
Notice that the value deduced in \eqref{eq:VH} can be written also as (see Annex 1 for the details)
\begin{itemize}
    \item[(a)] $\frac{\Gamma(H)\Gamma(1-H)}{\pi \Gamma(1+2H)}$,
    \item[(b)] $\frac{\Gamma(2-2H)\cos(\pi H)}{\pi H(1-2H)}$ (see \citep{DU1999}),
    \item[(c)] $\frac{1}{2H\sin(\pi H)\Gamma(2H)}=:\frac{A(H)}{\Gamma(H+1/2)^2}$, where $A(H)$ is the quantity considered for the covariance function by \citep{Lobodaetal2021} (cfr. equation (28)). This relation will be useful in section \ref{subsec:MPRE} to characterize the local behavior of a specific Multifractional Process with Random Exponent.
\end{itemize}
\end{remark}
\vspace{.5cm}
When fBm is sampled over a unit time interval ($T=1$), with $\text{Var}[B^H(1)]=V_H$, and $[0,1]$ is partitioned into $n$ subintervals, equation (\ref{eq:scalvar}) implies, for $n \to \infty$, $k=1,\ldots,n$
\begin{equation}
\lim_{n\to \infty} n^{2H-1} \sum_{k=1}^n (B^H_{k/n}-B^H_{(k-1)/n})^2 = V_H
\end{equation}
which entails
\begin{equation}\label{eq:sigmaH1}
\sigma_n = \text{StDev}(B^H_{k/n}-B^H_{(k-1)/n}) = \sqrt{V_H}n^{-H}.
\end{equation}
A generalization of (\ref{eq:sigmaH1}), which in turn extends  the rule of thumb commonly used to annualize volatility, will be used in Section \ref{sec:Application} to fit real data and relate volatility and regularity.\\

\begin{remark}
A fundamental relationship connects the value of the Hurst parameter and the regularity of the process trajectories. From a mathematical point of view, the degree of regularity of the graph of a function is quantified by its Hölder exponent, defined as follows.\\ 
Let \(Y = \{Y(t)\}_{t \in \mathbb{R}}\) be a stochastic process with continuous and nowhere differentiable paths. The pointwise Hölder exponent \(\alpha_Y(\tau)\) and the local Hölder exponent \(\tilde{\alpha}_Y(\tau)\) at a point \(\tau\) are defined as
\begin{equation}\label{eq:pointwise}
    \alpha_Y(\tau) := \sup \left\{ a \in [0, 1] : \limsup_{t \to \tau} \frac{|Y(t) - Y(\tau)|}{|t - \tau|^a} < +\infty \right\}, 
\end{equation}
\begin{equation}\label{eq:local}
    \tilde{\alpha}_Y(\tau) := \sup \left\{ \tilde{a} \in [0, 1] : \limsup_{(t', t'') \to (\tau, \tau)} \frac{|Y(t') - Y(t'')|}{|t' - t''|^{\tilde{a}}} < +\infty \right\}.
\end{equation}
The geometrical intuition of (\ref{eq:pointwise}) is as follows: function $Y$ has exponent $\alpha_Y(\tau)$ at $\tau$ if, for any positive $\epsilon$, there exists a neighborhood of $\tau$ such that the graph of $Y$ in the neighborhood is included in the envelope defined by $t \mapsto Y(\tau) - c|t-\tau|^{\alpha_Y(\tau)-\epsilon}$ and $t \mapsto Y(\tau)+c|t-\tau|^{\alpha_Y(\tau)+\epsilon}$ (Figure \ref{fig:Pointwise}). 
For certain classes of stochastic processes, including Gaussian processes, by virtue of zero-one law, there exists a non random quantity $a_Y(\tau)$ such that $\mathbb{P}(a_Y(\tau)=\alpha_Y(\tau))=1$ \citep{Ayache2013}. When $Y(t)$ is a semimartingale (e.g. Brownian motion), $\alpha_Y=\frac{1}{2}$; values different from $\frac{1}{2}$ describe non-Markovian processes, whose smoothness is too high $\left(\alpha_Y \in \left(\frac{1}{2},1\right)\right)$, or too low $\left(\alpha_Y \in \left(0,\frac{1}{2}\right)\right)$, to satisfy the martingale property. In particular, the quadratic variation of the process can be proven to be zero if $\alpha_Y>\frac{1}{2}$, and infinite if $\alpha_Y<\frac{1}{2}$ \citep{LVB2008, AyacheBouly2023} .
\begin{figure}
\centering
\captionsetup{margin=2.8cm}
\includegraphics[scale=.3, trim=0pt 0pt 0pt 0pt]{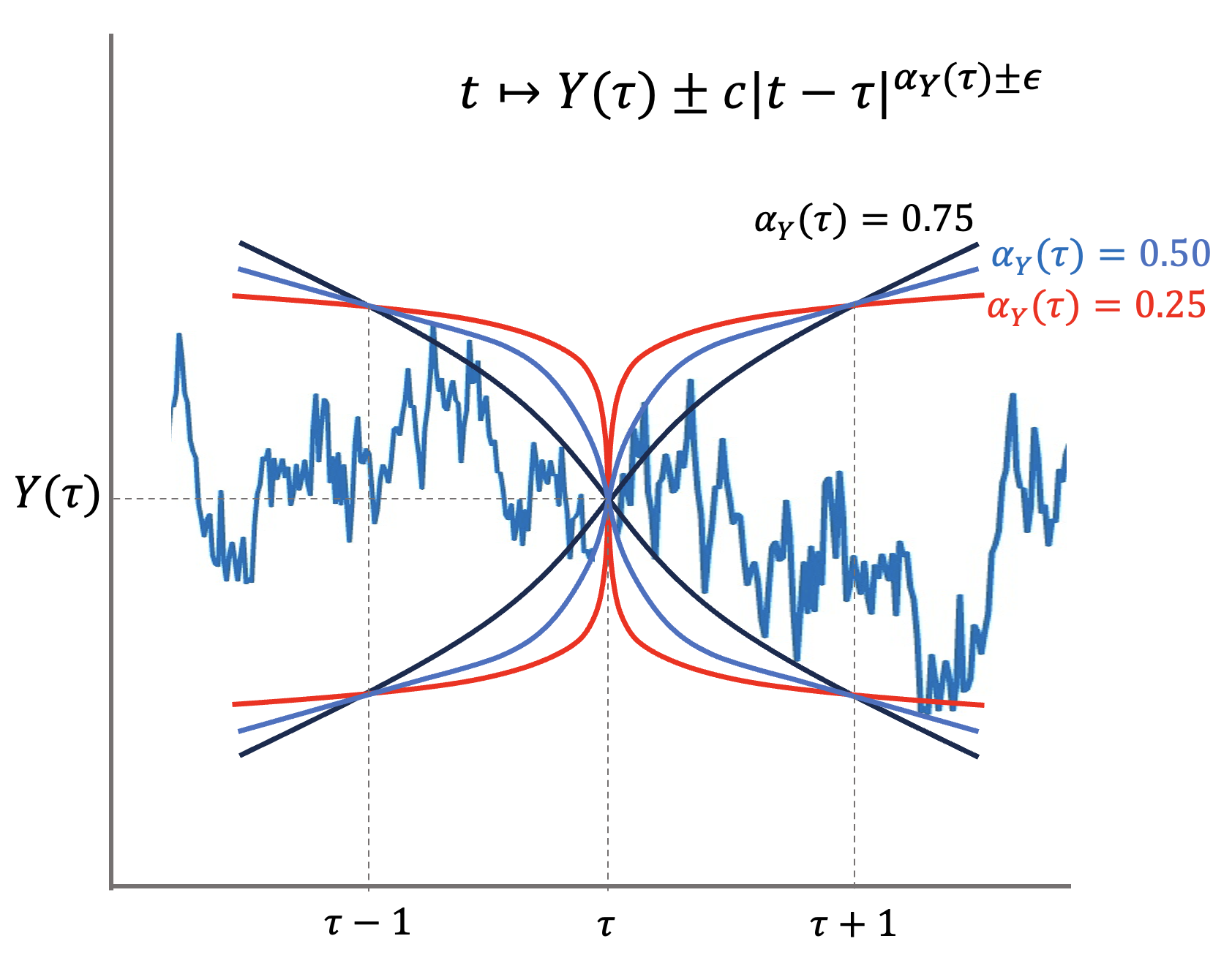}
\caption{Geometrical intuition of the pointwise H\"older exponent. For a given $\omega \in \Omega$, the closer to zero are $\alpha_Y(\tau,\omega)$ and $\tilde{\alpha}_Y(\tau,\omega)$, the rougher is the path $t \mapsto Y(t, \omega)$ in the vicinity of $\tau$; the closer to one are $\alpha_Y(\tau,\omega)$ and $\tilde{\alpha}_Y(\tau,\omega)$, the smoother is the path $t \mapsto Y(t, \omega)$ in the vicinity of $\tau$.}
\label{fig:Pointwise} 
\end{figure}
\\
\vspace{.25cm}\\
FBm's stationary increments $B^H(t + \tau) - B^H(t)$ are normally distributed with mean 0 and variance proportional to $\tau^{2H}$. It can be proved that its pointwise H\"older exponent $\alpha_{B_H}(t) = \tilde{\alpha}_{B_H}(t)= H$  almost surely for all $t$. Notice that its special case Brownian motion $B(t)=B^{1/2}(t)$ has Hölder exponent $\alpha_B(t) =\tilde{\alpha}_{B}(t)= \frac{1}{2}$ almost surely for all $t$.
Thus, the path roughness of fBm remains the same everywhere and one has, for all $\tau \in \mathbb{R}$, \citep{Xiao1997}
\begin{equation*}
    \mathbb{P}\left(\alpha_{B_H}(\tau)=\tilde{\alpha}_{B_H}(\tau)\right)=1.
\end{equation*}
Since a constant Hurst-Hölder exponent may be too limiting for many real-world applications (e.g. in the case of a stochastic volatility model), Péltier and Lévy Véhel \cite{PL1995} and Benassi et al. \cite{BJR1997} independently have built a more general Gaussian process with non-stationary increments and continuous paths named Multifractional Brownian Motion (MBM). It replaces the Hurst parameter $H$ of equation (\ref{eq:fbm}) by a deterministic continuous function $h:t \mapsto h(t)\in [\underline{H},\overline{H}] \subset (0,1)$. Function $h$ provides also the local path roughness of MBM, in the sense that for any point $t\in\mathbb{R}$ at which $h(t) < \tilde\alpha_h(t)$, it is almost surely $\alpha_{B_{h(t)}}(t) = h(t)$. In the next paragraph, we will introduce an even more general process (the Multifractional Process with Random Exponent) whose functional Hurst parameter is  no longer deterministic.\\
\end{remark}

\subsection{Multifractional Process with Random Exponent}\label{subsec:MPRE}
As noted in the preceding section, assuming a constant Hurst parameter is overly restrictive and unrealistic for many processes that exhibit regime changes or heteroskedasticity—features that are, by definition, characteristic of stochastic volatility models. To address the limit of a constant Hurst exponent, the fBm has been broadened in several directions\footnote{Generalizations encompass various models, such as the multifractional Brownian motion (mBm) \cite{PL1995,BJR1997} and its Generalized version (GmBm) \cite{AL2000}, the Multifractional Processes with Random Exponents (MPRE) \cite{AT2005}, the bifractional Brownian motion \cite{HV2003}, the mixed fBm \cite{Che2001}, the fractional Riesz-Bessel motion \cite{Anh1999}, the Multi-fractional Generalized Cauchy Process \cite{Li2020} (for a comprehensive overview, see \cite{Lim2015}).}, but one of the most comprehensive and flexible generalizations is provided by the \textit{Multifractional Processes with Random Exponent} (MPRE). Using random wavelet series, \citep{AT2005,AJT2007} introduced a first type of MPRE and shown that, under mild conditions, the roughness of the process at time $t$ can be prescribed by the value of $H(t)$, this justifying the name of Hurst-Hölder regularity. Unfortunately the MPRE discussed in \citep{AT2005, AJT2007} cannot be represented through the usual Itô integral; in order to avoid this drawback, a different type of MPRE was introduced later in \citep{AEH2018} and generalized recently in \citep{Lobodaetal2021}. In the following, we will refer to this process.\\
Given the filtered probability space $(\Omega, \pazccal{F}, (\pazccal{F}_s)_{s\in \mathbb{R}}, \mathbb{P})$, let $B =\{B(s)\}_{s\in \mathbb{R}}$ be a standard Brownian motion with respect to the filtration $(\pazccal{F}_s)$ and $g(t,s)$ be $\pazccal{F}_s$-adapted such that $g(t,s) = 0$ for $s > t$ with $\int_{-\infty}^t |g(t,s)|^2ds < \infty$ for all $t$.\\
The MPRE $K^H(t)$ is defined, for each $t\in \mathbb{R}^+$, as the Itô integral representation
\begin{equation} \label{eq:MPRE0}
K^H(t)=\int_{-\infty}^t g(t,s)dB(s),
\end{equation}
To fulfill desirable regularity properties, function $g(t,s)$ must adhere to the following criteria (for ease of reference, these criteria are labeled with the same letters as in \cite{Lobodaetal2021}):
\begin{description}
\item[(A)] \hspace{.1cm} Function $t \to g(t,s)$ is differentiable in $t>s$, for all $s$. There exist $\pazccal{F}_s$-adapted processes $H(s)$, $L(s)$ and $R(s)$, such that $H(s) \in (0,1)$, $L(s) >0$, $R(s) >1/2$ and for all $t \geq 0$,
\begin{flalign} \nonumber
\text{A1. } &|g(t,s)| \leq L(s)|t-s|^{H(s)-1/2}, \quad \forall s \in (t-1,t)&&\\ \nonumber
\text{A2. } &|\partial_t g(t,s)| \leq L(s) |t-s|^{H(s) - 3/2}, \quad \forall s \in (t-1,t)&&\\ \nonumber 
\text{A3. } &|\partial_t g(t,s)| \leq L(s) |t-s|^{-R(s)}, \quad \forall s \in (-\infty, t-1].&&\nonumber 
\end{flalign}
\item[(L-BC)] There exists a sequence of stopping times $\tau_n \to \infty$, real numbers $\underline{H}_n$, $\overline{H}_n$, $\overline{L}_n$, $\underline{R}_n$, and a sequence of continuous, increasing functions $\omega_n(h)$ with $\omega_n(0) = 0$, such that
\begin{flalign} \nonumber
\text{L-BC1. } &|H_{(t+h) \land \tau_n}-H_{t \land \tau_n}| \leq \omega_n(h), \quad \text{(continuity of $H(t)$)}&& \\ \nonumber
\text{L-BC2. } &0<\underline{H}_n \leq H_{t \land \tau_n} \leq \overline{H}_n < 1, \quad \text{(boundedness of $H$)}&&\\ \nonumber
\text{L-BC3. } &|L_{t \land \tau_n}| \leq \overline{L}_n, \quad \text{(local boundedness of $L$)}&&\\ \nonumber
\text{L-BC4. } &|R_{t \land \tau_n}| \geq \underline{R}_n > \frac{1}{2} \quad \text{(local boundedness of $R$).}&&\nonumber
\end{flalign}
\end{description}
Function $g(t,s) := \nu(s)[(t-s)_+^{H(s)-1/2}-(-s)_+^{H(s)-1/2}]$, where $\nu(s)$ is continuous and bounded deterministically, satisfies Assumptions (A) and (L-BC). Therefore, a special case of the general MPRE process (\ref{eq:MPRE0}) is 
\begin{equation} \label{eq:MPRE1}
K_1^H(t) = \int_{-\infty}^t \nu(s) \left[(t-s)_+^{H(s)-1/2}-(-s)_+^{H(s)-1/2}\right] dB_s,
\end{equation}
With respect to specification \eqref{eq:MPRE1} of the MPRE, it is worthwhile to observe that:
\begin{enumerate}
    \item since $g(t,s)$ satisfies all the conditions of Theorem 1.5 in \citep{AyacheBouly2022}, p.146, $\alpha_{K_1^H}(\tau)=H(\tau)$, that is the pointwise regularity of $K_1^H(\tau)$ is given by the Hurst-H\"older exponent at $\tau$;
    \item in the particular case where all the random variables $H(s)$ are equal to the same deterministic constant $H$, $K_1^H(t)$ reduces to the $B^H(t)$;
    \item assuming that $H(t)\in(0,1)$ and $\nu(t)>0$ are stationary adapted processes, with respect to representation \eqref{eq:fbm01} \citep{Lobodaetal2021} prove that
    \begin{equation}
        \mathbb{E}(K_1^H(t)K_1^H(s)) = \mathbb{E}\left[\nu(0)^2\frac{A(H(0))}{2}\left(|t|^{2H(0)}+|s|^{2H(0)}-|t-s|^{2H(0)}\right)\right]
    \end{equation}
    provided that the latter value is finite for all $t,s\in\mathbb{R}$.\\ 
    From this, it can easily be shown that the autocovariance function $\gamma(\Delta)$, $\Delta \in \mathbb{N}$, of the increment  process $(K_1^H(t+1)-K_1^H(t))$ is
    \begin{eqnarray}
    \gamma(\Delta)&=&\mathbb{E}\left[\left(K_1^H(t+1)-K_1^H(t)\right)\left(K_1^H(t+\Delta+1)-K_1^H(t+\Delta)\right)\right] \nonumber \\
        &=& \mathbb{E}\left[\nu(0)^2\frac{A(H(0))}{2}\left(|\Delta+1|^{2H(0)}-2|\Delta|^{2H(0)}+|\Delta-1|^{2H(0)}\right)\right] \nonumber \\
         &=& \mathbb{E}\left[\nu(0)^2\frac{\Gamma\left(H(0)+\frac{1}{2}\right)^2V_{H(0)}}{2}\left(|\Delta+1|^{2H(0)}-2|\Delta|^{2H(0)}+|\Delta-1|^{2H(0)}\right)\right] \nonumber
    \end{eqnarray} 
    where the last equality is justified by proof (c) in Remark \ref{rem:Remark1}.\\
    Of course, it is
    \begin{eqnarray} \label{eq:VarK}
        \text{Var}\left(K_1^H(t+1)-K_1^H(t)\right) &=&\mathbb{E}\left[\nu(0)^2\Gamma\left(H(0)+\frac{1}{2}\right)^2V_{H(0)}\right] \nonumber \\        &=&\mathbb{E}\left[\nu(0)^2\Gamma\left(H(0)+\frac{1}{2}\right)^2\frac{\Gamma(1-2H(0))}{\pi H(0)} \cos (\pi H(0))\right] \nonumber \\
        &=& \mathbb{E}\left[\nu(0)^2\frac{\Gamma\left(H(0)+\frac{1}{2}\right)\Gamma\left(1-H(0)\right)}{2^{2H(0)}H(0)\sqrt{\pi}}\right] \nonumber \\
        &=& \mathbb{E}\left[\nu(0)^2\frac{2^{1-4H(0)}\pi\Gamma(2H(0))}{H(0)\Gamma(H(0))^2\sin(\pi H(0))}\right]
    \end{eqnarray}
where the last equality -- deduced by applying the Euler's reflection identity and the Legendre's duplication formula (see Annex 2 for the proof) -- is more suitable for numerical evaluation, especially when $\Gamma(1-H)$ is hard to evaluate near $H=1$. Figure \ref{fig:VarIncr1} displays function $E(H)=\frac{2^{1-4H(0)}\pi\Gamma(2H(0))}{H(0)\Gamma(H(0))^2\sin(\pi H(0))}$.\\
\begin{figure}
\centering
\captionsetup{margin=3.33cm}
\includegraphics[scale=.125, trim=0pt 0pt 0pt 0pt]{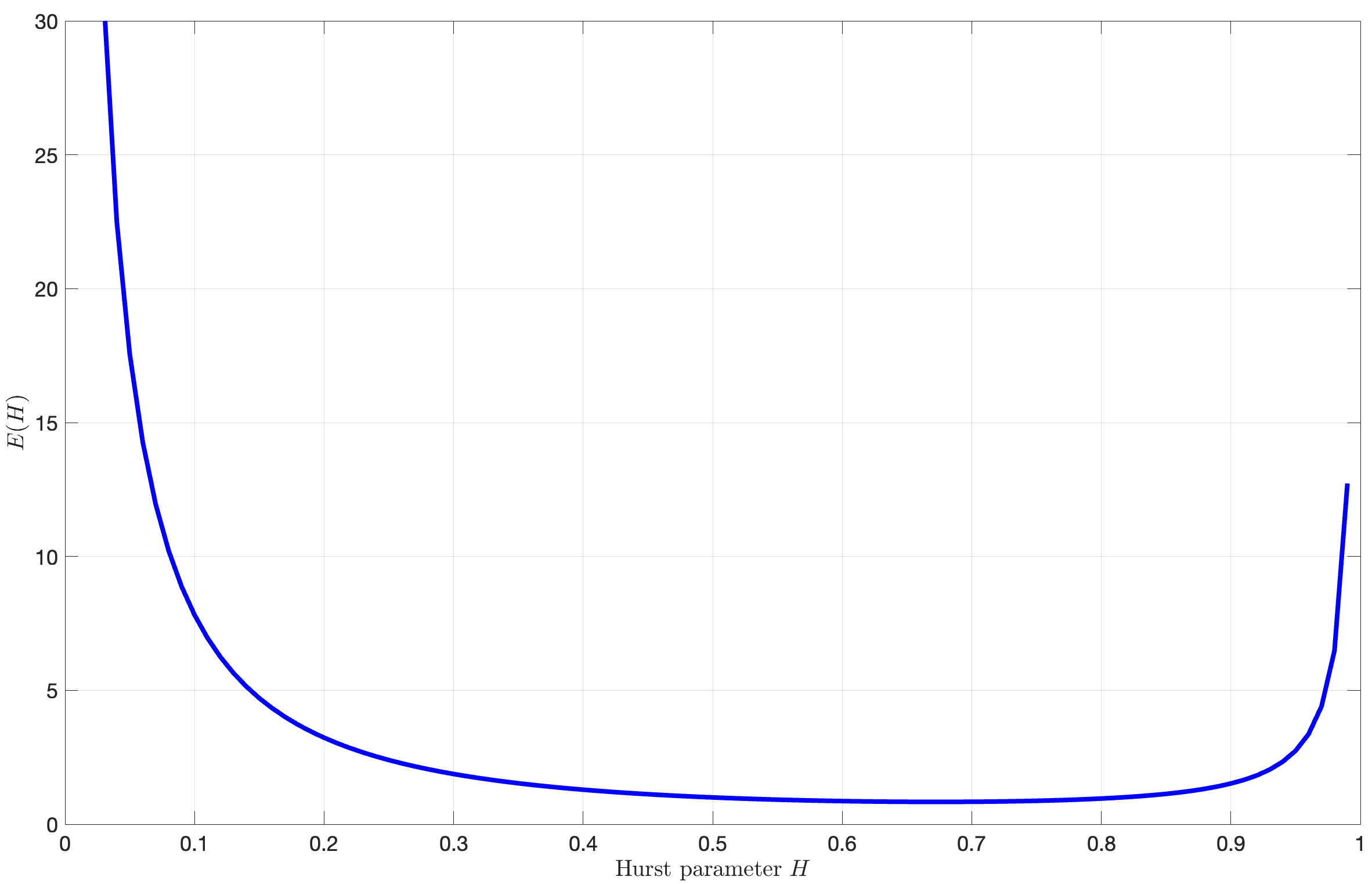}
\caption{Function $E(H) = \frac{2^{1-4H(0)}\pi\Gamma(2H(0))}{H(0)\Gamma(H(0))^2\sin(\pi H(0))}$.}
\label{fig:VarIncr1} 
\end{figure}

Net of function $\nu^2$, one can calculate \eqref{eq:VarK} once the dynamics of $H$ is specified. For example one can model $H(t)$ by an eventually fractional Ornstein-Uhlenbeck process. In this case, assuming $H(t) \sim \mathcal{N}\left(\frac{1}{2},\sigma^2_H\right)$, where $\sigma_H^2$ is taken sufficiently small to make negligible the probability that the Hurst-Hölder exponent lies outside the interval $(0,1)$ (e.g. $\mathbb{P}(0<H<1)\geq 0.999$ implies $\sigma_H\leq 0.023$), an approximation of \eqref{eq:VarK} for a fixed time $t$ can be obtained by a second-order Taylor expansion of $E(H)$ around $H=\frac{1}{2}$. 
Since $\mathbb{E}(H)=\frac{1}{2}$ and $\mathbb{E}\left[\left(H-\frac{1}{2}\right)^2\right] = \sigma^2_H$, one has 
\begin{equation*}
    \mathbb{E}(E(H)) \approx E\left(\frac{1}{2}\right)+\frac{\sigma_H^2}{2}E''\left(\frac{1}{2}\right)
\end{equation*}
while the first-order term vanishes due to symmetry. Thus (see steps in Annex 3)
\begin{equation*}
    \mathbb{E}(E(H)) \approx 1+\left(4+\frac{\pi^2}{3}\right)\sigma^2_H
\end{equation*}
Table \ref{tab:E(H)} provides the values of variance \eqref{eq:VarK}, net of function $\nu$, for different values of variance $\sigma_H^2$ of a normally distributed Hurst-Hölder exponent.

\begin{table}[ht]
\centering
\caption{Expected values and probabilities for different variances of $H \sim \mathcal{N}(1/2, \sigma_H^2)$}
\label{tab:E(H)}
\small
\begin{tabular}{c c c}
\toprule
{Variance $\sigma_H^2$} & {$\mathbb{P}[H \notin (0,1)]$} & {$\mathbb{E}[E(H)]$} \\
\midrule
0.0230 & $9.7757\times10^{-04}$ & 1.1677 \\
0.0200 & $4.0695\times10^{-04}$ & 1.1458 \\
0.0175 & $1.5705\times10^{-04}$ & 1.1276 \\
0.0150 & $4.4557\times10^{-05}$ & 1.1093 \\
0.0125 & $7.7442\times10^{-06}$ & 1.0911 \\
0.0100 & $5.7330\times10^{-07}$ & 1.0729 \\
0.0050 & $1.5375\times10^{-12}$ & 1.0364 \\
0.0010 & $2.5968\times10^{-56}$ & 1.0073 \\
\bottomrule
\end{tabular}
\end{table}
\item more generally, without requiring the stationarity of $H$ and $\nu$, one can establish a rescaling limit property -- referred to as the \textit{Local Asymptotic Self-Similarity} (LASS) -- showing that, for each fixed $t$, as $h \to 0$,
\begin{eqnarray} \label{eq:LASS1}
h^{-H(t)}\left(K_1^H(t+hr)-K_1^H(t)\right) &\implies& \nu(t) \int_{-\infty}^r \left[(r-s)_+^{H(t)-1/2}-(-s)_+^{H(t)-1/2}\right]d\tilde{B}(s) \nonumber \\
&\sim& \nu(t)\Gamma\left(H(t)+\frac{1}{2}\right)B^{H(t)}(r) \quad  \quad \text{[by \eqref{eq:fbm}] }
\end{eqnarray}
where $\tilde{B}(s)$ is a Brownian motion independent on $H(t)$ (see  \cite{Lobodaetal2021} and \citep{Reed1995}).\\
This means that
\begin{equation}
    K_1^H(t+hr)-K_1^H(t) \sim  \nu(t)\Gamma\left(H(t)+\frac{1}{2}\right)(hr)^{H(t)}B^{H(t)}(1) \quad \text{as }h \to 0.
\end{equation}
Comparison of equations \eqref{eq:LASS1} and \eqref{eq:fbm} shows that, in a neighborhood of any point $t$, $K_1^H(t)$ exhibits the behavior of an fBm with Hurst exponent $H(t)$, that is at any point $t$ along the sample paths of the MPRE there exists an fBm with parameter $H(t)$ tangent to $K_1^H(t)$\footnote{A tangent process of a stochastic process $X(t)$ at a point $t$ is defined to be the limit of a sequence of scaled enlargements of $X(t)$ about $t$. The tangent process reflects the local structure of $X(t)$ and, because of its locally stationary increments, it allows to manage the eventual non stationarity of $X(t)$ \cite{Falconer2002}.}. This is a fundamental property, since it implies that equation (\ref{eq:sigmaH1}) holds locally also for the MPRE.
\end{enumerate}

\section{Risk and volatility through the Hurst-H\"{o}lder exponent} \label{sec:RiskVol}
As discussed above, the Hurst–Hölder regularity quantifies the local pathwise regularity of stochastic processes and serves as a measure of deviation from semimartingale behavior, with the critical threshold at $H=\frac{1}{2}$ corresponding to martingale dynamics. As such, it provides a rigorous theoretical foundation for modeling volatility through the Hurst–Hölder exponent, particularly in frameworks where roughness is empirically observed and martingale properties are relaxed. This has several advantages:
\begin{itemize}[leftmargin=*]\setlength\itemsep{0.25cm}
    \item[a)] the Hurst-H\"older exponent is responsive to autocorrelation regardless of the scale parameter, while volatility is not. With appropriate scale parameters, data with high or low correlation can exhibit identical volatility, which is unreasonable if volatility is meant to gauge financial risk. On the other hand, data with varying levels of autocorrelation have different Hurst-H\"older parameters, since the intensity of autocorrelation influences the smoothness of process trajectories. 
    \item[b)] Volatility is a comparative measure rather than an absolute one. It just indicates whether a market or an asset currently exhibits more or less variability compared to previous periods, but determining the \say{\textit{optimal}} or \say{\textit{fair}} level of volatility, that is the level of volatility which is consistent with an efficient market, is not within its capacity. In contrast, the Hurst-H\"older parameter lies within the range of $(0,1)$ and equals $1/2$ only when the process aligns with a Brownian motion, which serves as the epitome of the EMH. As outlined in \cite{BPP2015}, the magnitude of inefficiency characterizing the market at any given time $t$ can be captured by the measure $|H(t)-1/2|\in(0,1/2)$ which provides information about the degree of predictability of future values. 
    \item[c)] Given that, once they have deviated from it, markets tend to \textit{naturally} return to the equilibrium $H(t)=1/2$, the distance $|H(t)-1/2|$ provides a non-trivial indicator for the timing of buying/selling. Indeed, it is reasonable to expect that the dynamics of $H(t)$ exhibit a fluctuating trend around the level $1/2$, with the rate of return to this value being higher the wider the deviation. This mechanism constitutes a stochastic formalization and a theoretically grounded explanation of the well-known adage among traders \say{\textit{What goes up, must come down}}, and a parsimonious conceptual explanation for the findings by \cite{DeBondt1995}; as is well-known, they observe systematic price reversals for stocks that experience extreme long-term gains or losses to the extent that past losers significantly outperform past winners. 
    \item[d)] When financial prices are represented by processes that exhibit local behavior akin to a fBm (as is the case with MPRE), the relationship between volatility and the Hurst-H\"older exponent at time $t$ can be expressed through equations (\ref{eq:sigmaH1}) or (\ref{eq:sigmaH2}). As a result, there is no loss of information when employing the Hurst-H\"older parameter in lieu of volatility. In recent years, numerous empirical estimations (refer, for instance, to \cite{Cajueiro2004,BP2011,Peng2018,Garcin2022,Mattera2022}) have provided evidence indicating that the time-varying Hurst-H\"older exponent of financial time series displays stationarity, mean-reversion, and an approximately normal distribution around the value $1/2$. These observed patterns are also consistent with the assumption that a properly calibrated fractional Ornstein-Uhlenbeck process can be used to model the behavior of $H(t)$ (see Remark \ref{Rem:2}).
\end{itemize}
Table \ref{tab:FinIn} summarizes how the Hurst-H\"older exponent and the martingale condition are related and provides a  financial characterization of this link \citep{BP2018}. In contrast to volatility, the Hurst-H\"older exponent provides a holistic evaluation of market dynamics, addressing both the magnitude (\textit{"how much"}) and character (\textit{"how"}) of price variability. It offers insights into the deviation from equilibrium expressed by the value $H(t)=1/2$, which serves as the benchmark for a semi-martingale. The pointwise Hurst-H\"older exponent serves as a descriptor of the dominant dynamics at a specific point in time, discerning among \textit{momentum market} (linked to bullish phases or speculative bubbles), \textit{sideways market} (indicative of efficiency characterized by limited bull or bear trends), and \textit{reversal market} (resulting from rapid buy-and-sell activities, frequently following substantial price adjustments or periods of uncertainty).


\begin{table}[ht]
\caption{Financial interpretation of $H(t)$}
\label{tab:FinIn}
\footnotesize
\centering
\renewcommand{\arraystretch}{1.3}
\begin{tabularx}{\textwidth}{>{\bfseries}c | X | X | X}
\toprule
$H(t)$ & Stochastic properties & Behavioral interpretation & Market implications \\
\midrule
$> \tfrac{1}{2}$ & 
Persistence, smooth paths, $[X](t) = 0$ &
New information confirms existing positions &
\textit{Unfair low} volatility, momentum, positive inefficiency, overconfidence, underreaction \\
\midrule
$= \tfrac{1}{2}$ & 
Independence, martingale behavior, $[X](t) = t$ (Brownian)&
Information fully incorporated into prices &
\textit{Fair} volatility, sideways market, informational efficiency \\
\midrule
$< \tfrac{1}{2}$ & 
Mean-reversion, rough paths, $[X](t) = \infty$ &
New information disrupts existing positions &
\textit{Unfair high} volatility, reversals, negative inefficiency, overreaction \\
\bottomrule
\end{tabularx}
\normalsize
\end{table}

This interpretation suggests that the apparently conflicting paradigms of Rationality and Behavioral Finance can coalesce within a comprehensive framework of bounded rationality, providing a more nuanced understanding of market dynamics. Within this framework, the pointwise Hurst-H\"older exponent explicitly identifies \textit{when} rationality transitions to irrationality. This transition may manifest as overconfidence, bullish trends, or speculative bubbles, or alternatively as overreaction, reversals, and panic-selling.

\section{Estimation of H\texorpdfstring{\textsubscript{t}}{t}}\label{sec:Hestimation}
The estimation of the Hurst-H\"older exponent has been widely studied in the literature, with variation statistics being a common approach (see, e.g., \cite{IL1997}, \cite{KW1997}, \cite{BBCI2000}, \cite{Coeur2001}, and \cite{Coeur2005}). As the estimation problem lies beyond the scope of this contribution, below we briefly outline the key steps of the procedure adopted in Section~\ref{sec:Application}, referring the reader to \cite{Garcin2017}, \cite{AngeliniBianchi2023} (Section 3), and \cite{Frezza2023} (Section 3) for a comprehensive exposition.\\

Consider a discretized sample \(\left(X_{t}, t \in [\![1,n]\!]\right)\) of a multifractional process over the interval \([0,1]\), where \(C^2\) denotes the unknown variance at unit time. Due to its local asymptotic similarity to fractional Brownian motion (fBm), the increments of \(X_t\) are approximately normally distributed in sufficiently small neighborhoods. For a small, even window size \(\delta \ll n\), where the regularity is assumed constant, the \(k\)-th absolute moment of the increments can be approximated by a normal distribution with a variance scaling law dependent on the Hurst-H\"older exponent \(H_t\).

Focusing on the second moment (\(k=2\)), the expected value of the moment estimator \(M_t^{(2)}\) follows a power-law relation with \(H_t\). To eliminate the unknown scale parameter \(C\), an auxiliary estimator \(M_t^{'(2)}\) is introduced, leading to the unbiased estimator proposed by \cite{Garcin2017}:
\[
\hat{H}_t^{2,\delta,n} = \frac{1}{2}\log_2 \frac{M_t^{'(2)}}{M_t^{(2)}}.
\]
\cite{PiaBiaPal2018} combine this with a biased estimator \(\hat{H}^{\delta,n,C^*}_t\), which converges at a rate of \(\pazccal{O}(\delta^{-1/2}(\log n)^{-1})\). As shown by Proposition 1 in \citep{AngeliniBianchi2023}, from this combination it is possible to derive the unbiased estimator $\hat{H}_t^{(\delta)}$. \\

Thus, equation (\ref{eq:sigmaH1}) locally holds for the MPRE in the form
\begin{equation}\label{eq:sigmaH2}
    \sigma_{t,n}=\sqrt{V_{\hat{H}_t^{2,\delta,n}}}\,\, n^{-\hat{H}_t^{2,\delta,n}}.
\end{equation}

\begin{remark}\label{Rem:2}
The log-transform of (\ref{eq:sigmaH2}) makes it explicit the relationship that can be established between the log-volatility $\sigma_{t,n}$ and the Hurst-H\"older exponent $H_t$ estimated through $\hat{H}_t^{2,\delta,n}$:
\begin{equation}\label{eq:linvol}
\ln \sigma_{t,n}=\frac{1}{2}\ln V_{\hat{H}_t^{2,\delta,n}}-\hat{H}_t^{2,\delta,n} \ln n.
\end{equation}
When the log-volatility is modeled by a fractional Ornstein-Uhlenbeck process, as for the Fractional Stochastic Volatility Model (FSV), equation (\ref{eq:linvol}) indicates that also the Hurst-H\"older exponent follows a Ornstein-Uhlenbeck process with parameters which are linear transforms of those used to model the log-volatility \citep{AngeliniBianchi2023}.
\end{remark}

Under the martingale hypothesis ($H_t = 1/2$), the estimator follows a normal distribution with mean $\frac{1}{2}$ and variance $\frac{1}{2\delta \log^2(n-1)}$ \cite{BPP2013}; this allows to build a confidence interval to test whether the martingale conditions holds., enabling the construction of confidence intervals for hypothesis testing. In Section \ref{sec:Application}, all estimates are computed using a rolling window of \(\delta = 20\) trading days.

\section{Real data analysis} \label{sec:Application}
\begin{figure}[H]

\centering
\begin{minipage}{0.485\textwidth}
  \includegraphics[width=\linewidth]{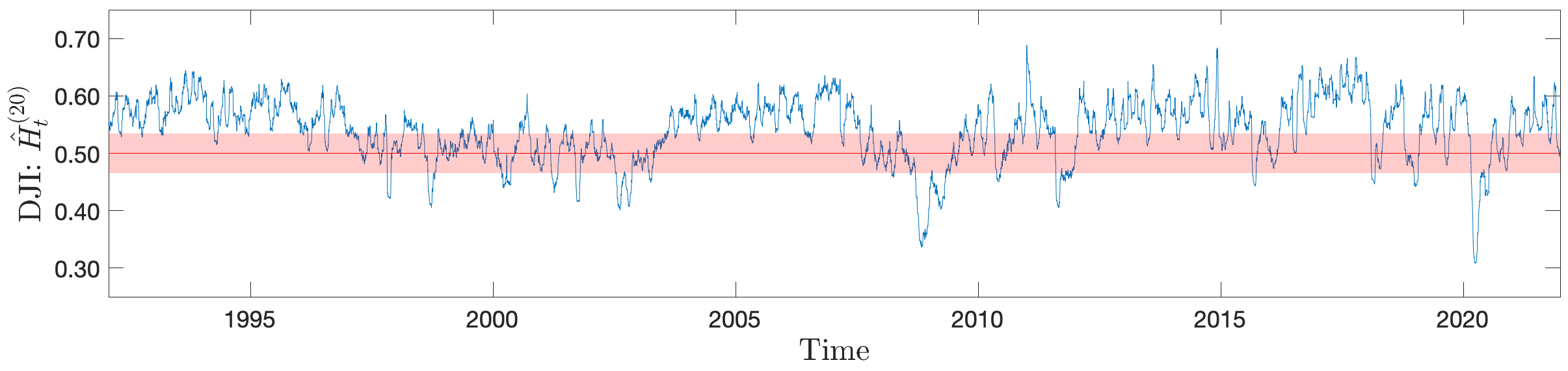}
\end{minipage}
\begin{minipage}{0.485\textwidth}
  \includegraphics[width=\linewidth]{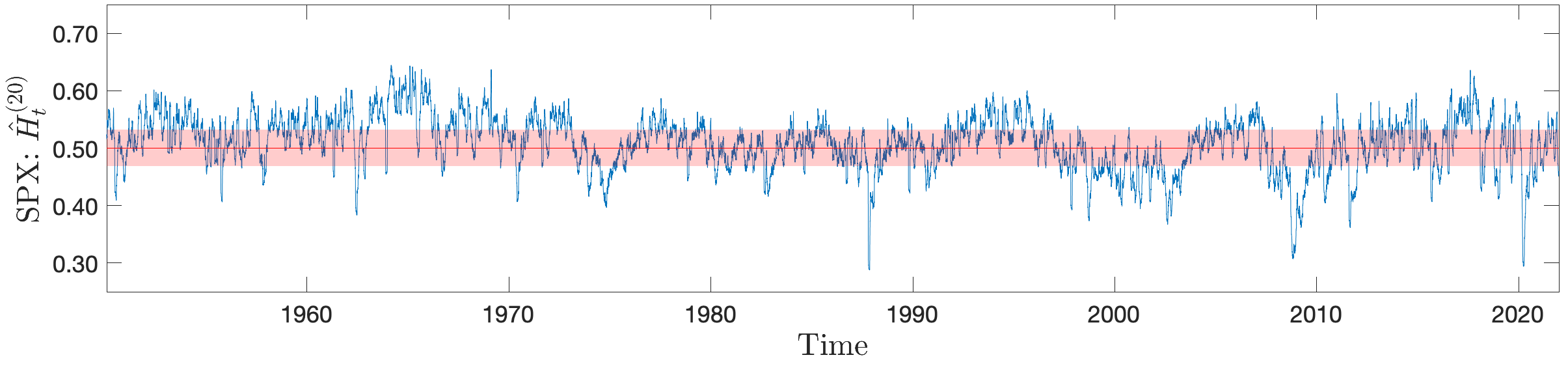}
\end{minipage}
\begin{minipage}{0.485\textwidth}
  \includegraphics[width=\linewidth]{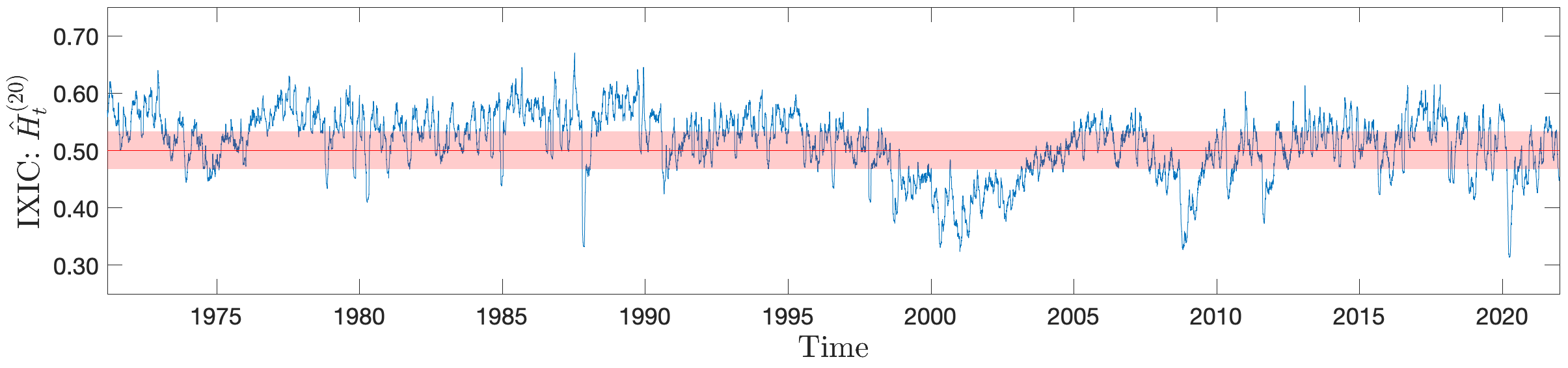}
\end{minipage}
\begin{minipage}{0.485\textwidth}
  \includegraphics[width=\linewidth]{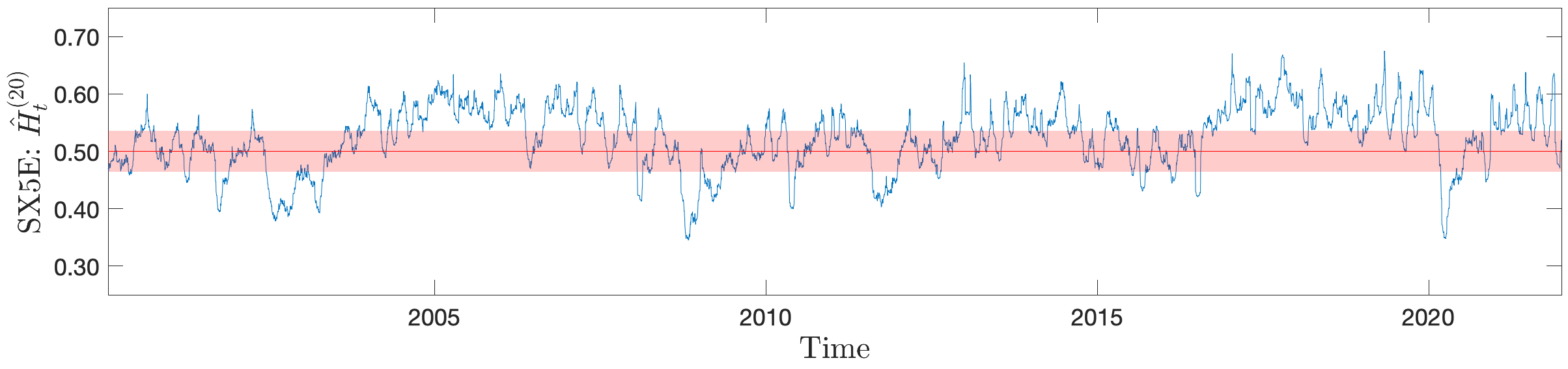}
\end{minipage}
\begin{minipage}{0.485\textwidth}
  \includegraphics[width=\linewidth]{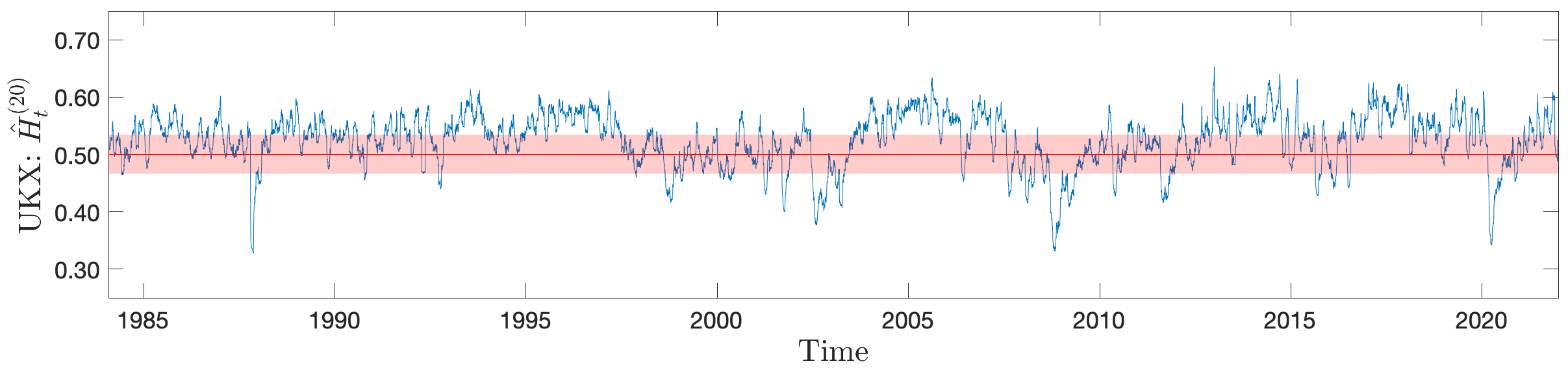}
\end{minipage}
\begin{minipage}{0.485\textwidth}
  \includegraphics[width=\linewidth]{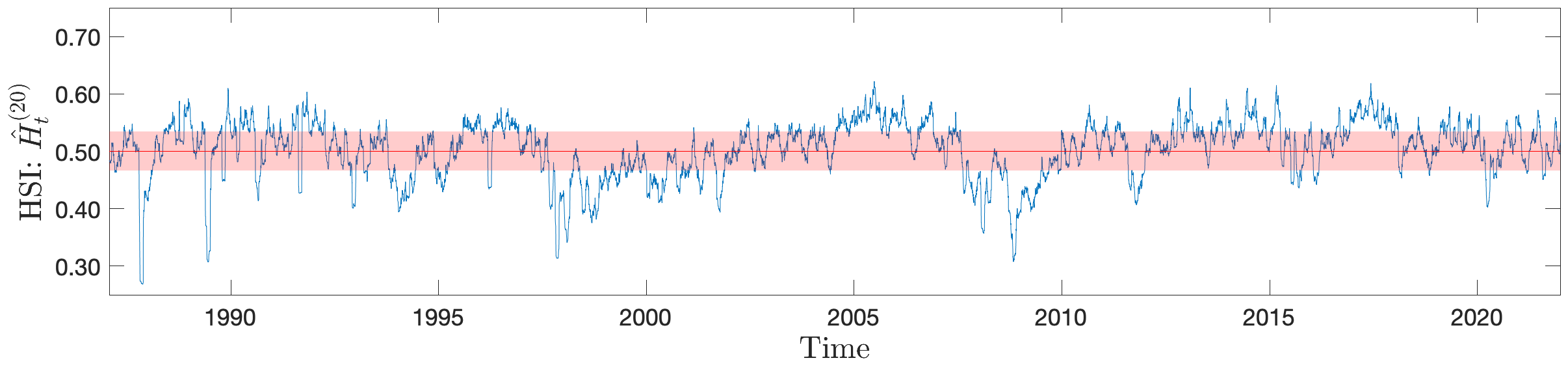}
\end{minipage}
\begin{minipage}{0.485\textwidth}
  \includegraphics[width=\linewidth]{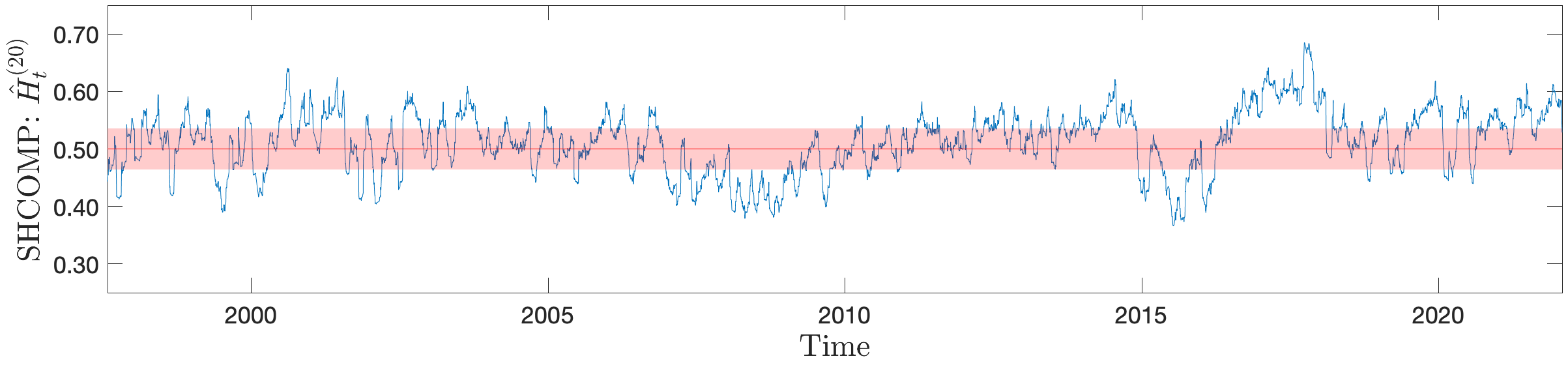}
\end{minipage}
\begin{minipage}{0.485\textwidth}
  \includegraphics[width=\linewidth]{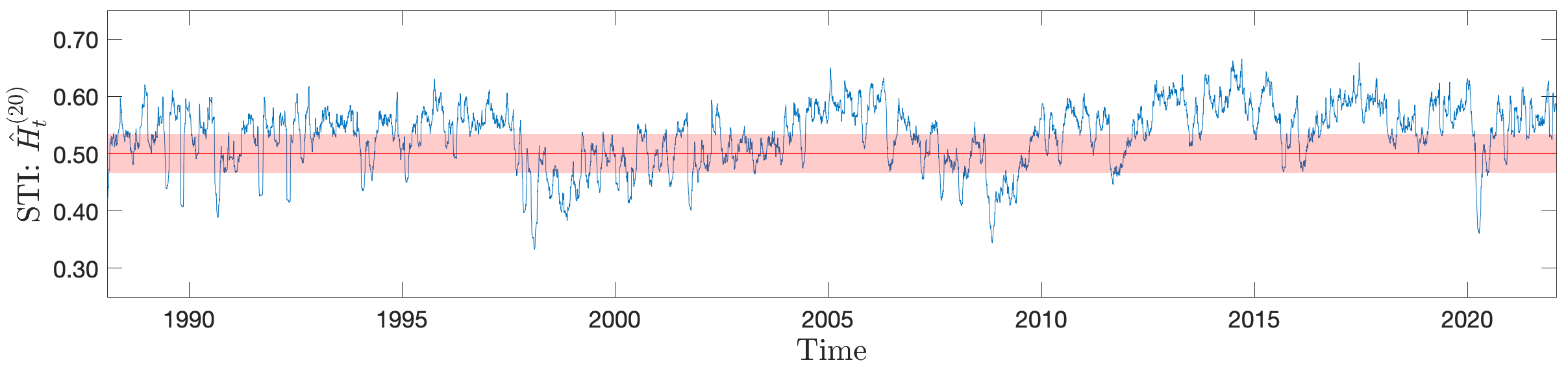}
\end{minipage}
\begin{minipage}{0.485\textwidth}
  \includegraphics[width=\linewidth]{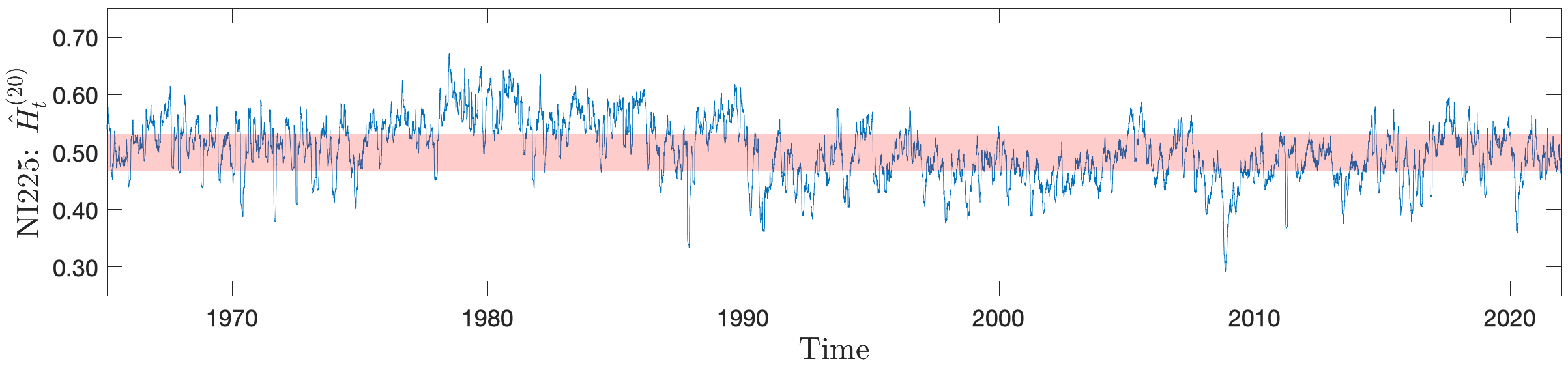}
\end{minipage}
\begin{minipage}{0.485\textwidth}
  \centering
  \captionsetup{labelformat=empty}
  \caption{%
    \parbox{0.90\linewidth}{
    \vspace{.1cm}
      Figure 5: Estimated pointwise Hurst exponents using a sliding window of $20$ trading days. 
      The light red regions refer to the $95\%$ confidence interval with respect to $H_t=\frac{1}{2}$.
    }
    \label{fig:Hestim0}
  }
\end{minipage}
\end{figure}

In this section, we present the results of an analysis estimating the Hurst exponent for nine stock market indexes over different time periods 
ranging from 1950 to the end of 2021: Dow Jones Industrial Average (DJI, USA), Standard \& Poor 500 (SPX, USA), Nasdaq Composite (IXIC, USA), Eurostoxx50 (SX5E, Europe), Footsie 100 (UKX, United Kingdom), Hang Seng (HSI, Hong Kong), SSE Composite (SHCOMP, China), Straits Times (STI, Singapore), Nikkei 225 (NI225, Japan). 
The extensive range of samples implies that the properties of the Hurst parameter remain consistent regardless of time, market, or the size of the time series. The analysis findings are outlined in Table \ref{tab:estim1} and illustrated in Figure \ref{fig:Hestim0}.\\
As expected, all the series exhibit mean-reverting behavior around the value $H_t=1/2$. For eight out of nine indexes, the average value falls within the 95\% confidence interval. As explained in the preceding section, the value $H_t=1/2$ signifies the absence of arbitrage opportunities, ensured by the martingale behavior of the price sequence. However, the frequency and duration of deviations of the Hurst-H\"older exponent from $1/2$ in all series are too frequent and prolonged to simplistically assert that arbitrage opportunities are consistently eliminated \textit{instantaneously} by the self-adjusting market mechanisms. Moreover, the time-varying nature of $H_t$ indicates that the quadratic variation depends on the observed time interval, necessitating a model for the dynamics of $H_t$ to determine the global probabilistic properties of the price process. 
In all cases, the distributions of the estimated values closely approximate the normal distribution. They generally exhibit kurtosis values slightly above 3 (ranging between $2.820$ and $4.636$) and moderate negative skewness, which ranges between $-0.969$ and $-0.123$ for all indexes. Notably, SX5E, SHCOMP, and NI225 are approximately symmetric, while the others display moderate negative skewness. The left skewness indicates the frequency of large negative variations in the parameter relative to positive variations. The former typically occur during market crashes and align with the asymmetric impact that new information has on market participants. It is improbable that new pieces of information can abruptly bolster market confidence in a trend, whereas it is more likely to destabilize the market. Confidence reinforcement is inherently a gradual process (reflected in small positive changes in $H_t$), whereas confidence erosion often constitutes an instantaneous shock when new information contradicts past beliefs or heightens uncertainty about the future. The table also presents the results of the augmented Dickey-Fuller test (ADF), which evaluates the null hypothesis of a unit root in the series of the estimated $H_t$. The alternative hypothesis is trend-stationarity. The ADF statistic takes a negative value; the smaller the value, the stronger the rejection of the hypothesis that there is a unit root at a given confidence level. Since the computed $p$-values are the smallest obtainable, the unit root is rejected for all series at every confidence level. Therefore, we conclude that the estimated sequences are trend-stationary.

\begin{table}[ht]
\centering
\caption{Main statistics of the estimated $H_t$}
\resizebox{\textwidth}{!}{ 
\begin{tabular}{lccccccccc}
\toprule
Index & DJI & SPX & IXIC & SX5E & UKX & HSI & SHCOMP & STI & NI225 \\
\midrule
Start date & 1992-01-02 & 1950-01-03 & 1971-02-05 & 2000-01-03 & 1984-01-03 & 1986-12-31 & 1997-07-02 & 1987-12-28 & 1965-01-05 \\
End date   & 2021-12-28 & 2021-12-28 & 2021-12-28 & 2021-12-31 & 2021-12-29 & 2021-12-29 & 2022-01-28 & 2022-01-28 & 2021-12-29 \\
\# Obs ($N$) & 7,555 & 18,101 & 12,470 & 5,730 & 9,599 & 8,612 & 5,955 & 8,409 & 13,797 \\
\midrule
Mean        & 0.540 & 0.508 & 0.512 & 0.524 & 0.524 & 0.502 & 0.512 & 0.533 & 0.502 \\
Range       & 0.308–0.688 & 0.288–0.644 & 0.314–0.670 & 0.345–0.675 & 0.329–0.652 & 0.269–0.622 & 0.366–0.685 & 0.333–0.665 & 0.292–0.672 \\
95\% C.I.   & (0.465, 0.535) & (0.468, 0.532) & (0.467, 0.533) & (0.462, 0.538) & (0.466, 0.534) & (0.466, 0.534) & (0.464, 0.536) & (0.466, 0.534) & (0.468, 0.533) \\
St. Dev.    & \phantom{-}0.054 & \phantom{-}0.047 & \phantom{-}0.055 & \phantom{-}0.055 & \phantom{-}0.046 & \phantom{-}0.0500 & \phantom{-}0.054 & \phantom{-}0.054 & \phantom{-}0.052 \\
Kurtosis    & \phantom{-}4.004 & \phantom{-}4.244 & \phantom{-}3.434 & \phantom{-}3.111 & \phantom{-}4.213 & \phantom{-}4.636 & \phantom{-}2.820 & \phantom{-}3.232 & \phantom{-}3.063 \\
Skew        & -0.714 & -0.600 & -0.690 & -0.437 & -0.832 & -0.969 & -0.203 & -0.630 & -0.123 \\
ADF stat    & -6.032 & \hspace{-0.5em}-10.223 & -7.697 & -5.690 & -7.358 & -7.312 & -6.069 & -7.053 & -9.112 \\
Critical v. & -3.413 & -3.412 & -3.412 & -3.413 & -3.412 & -3.413 & -3.413 & -3.413 & -3.412 \\
\bottomrule
\end{tabular}
}
\label{tab:estim1}
\end{table}

\normalsize

 %

\begin{table}
\centering
\caption{Statistics of the fit}
\small{Fit: $a + N^{-bH}\left( \frac{\Gamma(bH)\Gamma(1-bH)}{\pi \Gamma(1+2bH)} \right)^{1/2}$, where $N$ is the sample size.\\
\footnotesize\textit{Note:} Estimated values of $a$ and pertaining confidence bounds are multiplied by $10^4$.}
\resizebox{\textwidth}{!}{
\label{tab:FitStat}
\vspace{0.5em}
\renewcommand{\arraystretch}{1.3}
\small
\begin{tabular}{l c c c c c}
\toprule
\textbf{Index} & \textbf{R-squared} & \textbf{SSE} & \textbf{RMSE} & \textbf{Estimated $a$ (95\% CI)} & \textbf{Estimated $b$ (95\% CI)} \\
\midrule
DJI    & 0.9883 & 0.003128 & 0.0006443 & 8.114 (7.806, 8.423) & 1.010 (1.010, 1.011) \\
SPX    & 0.9842 & 0.007478 & 0.0006431 & 8.379 (8.175, 8.584) & 1.014 (1.013, 1.014) \\
IXIC   & 0.9868 & 0.007875 & 0.0007954 & 7.712 (7.425, 7.998) & 1.009 (1.009, 1.010) \\
SX5E   & 0.9871 & 0.003599 & 0.0007939 & 11.21 (10.73, 11.69) & 1.017 (1.016, 1.018) \\
UKX    & 0.9846 & 0.004023 & 0.0006481 & 10.89 (10.59, 11.19) & 1.030 (1.029, 1.030) \\
HSI    & 0.9762 & 0.015050 & 0.0013230 & 15.97 (15.36, 16.58) & 1.006 (1.005, 1.007) \\
SHCOMP & 0.9811 & 0.005099 & 0.0009270 & 10.95 (10.37, 11.54) & 1.018 (1.017, 1.019) \\
STI    & 0.9822 & 0.005151 & 0.0007836 & 7.931 (7.560, 8.311) & 1.004 (1.003, 1.005) \\
NI225  & 0.9848 & 0.007835 & 0.0007541 & 6.088 (5.787, 6.388) & 0.986 (0.986, 0.987) \\
\bottomrule
\end{tabular}
}
\normalsize
\end{table}

\normalsize
\begin{figure}
 \centering
\hspace{-.5cm}\includegraphics[scale=.15, trim=0pt 0pt 100pt 0pt]{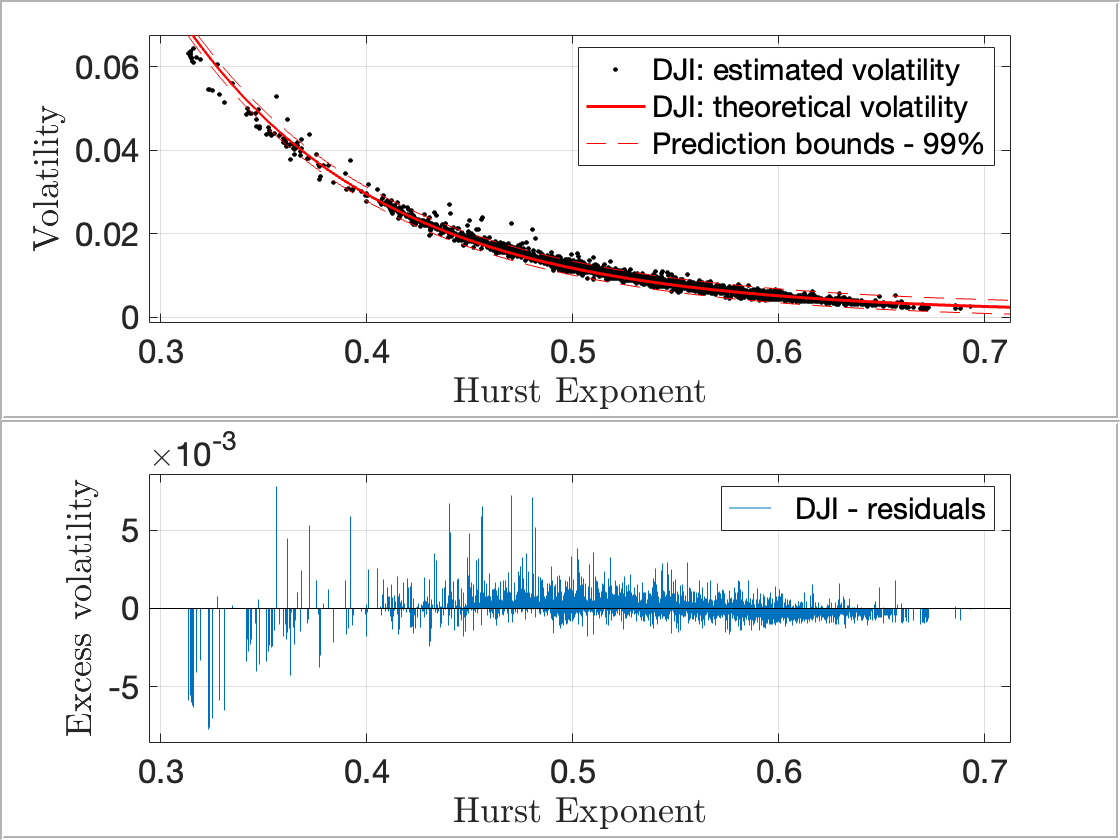}
\includegraphics[scale=.15, trim=0pt 0pt 100pt 0pt]{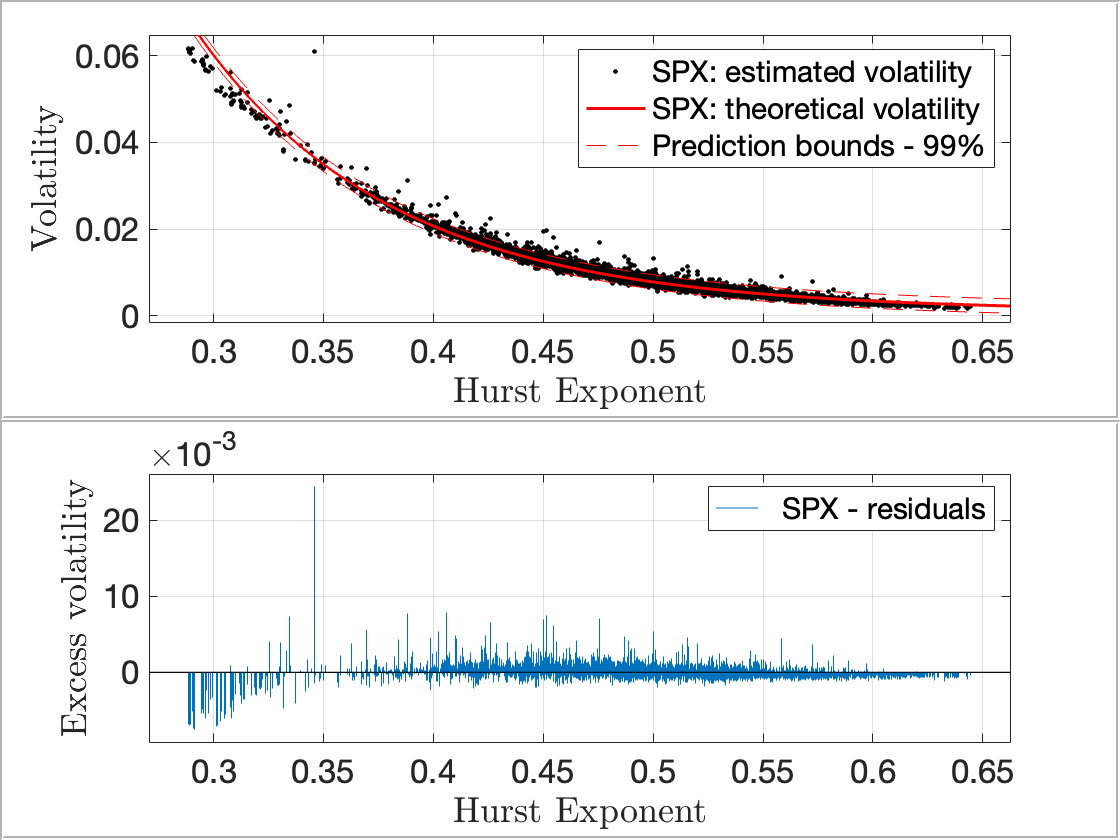}
\includegraphics[scale=.15, trim=0pt 0pt 100pt 0pt]{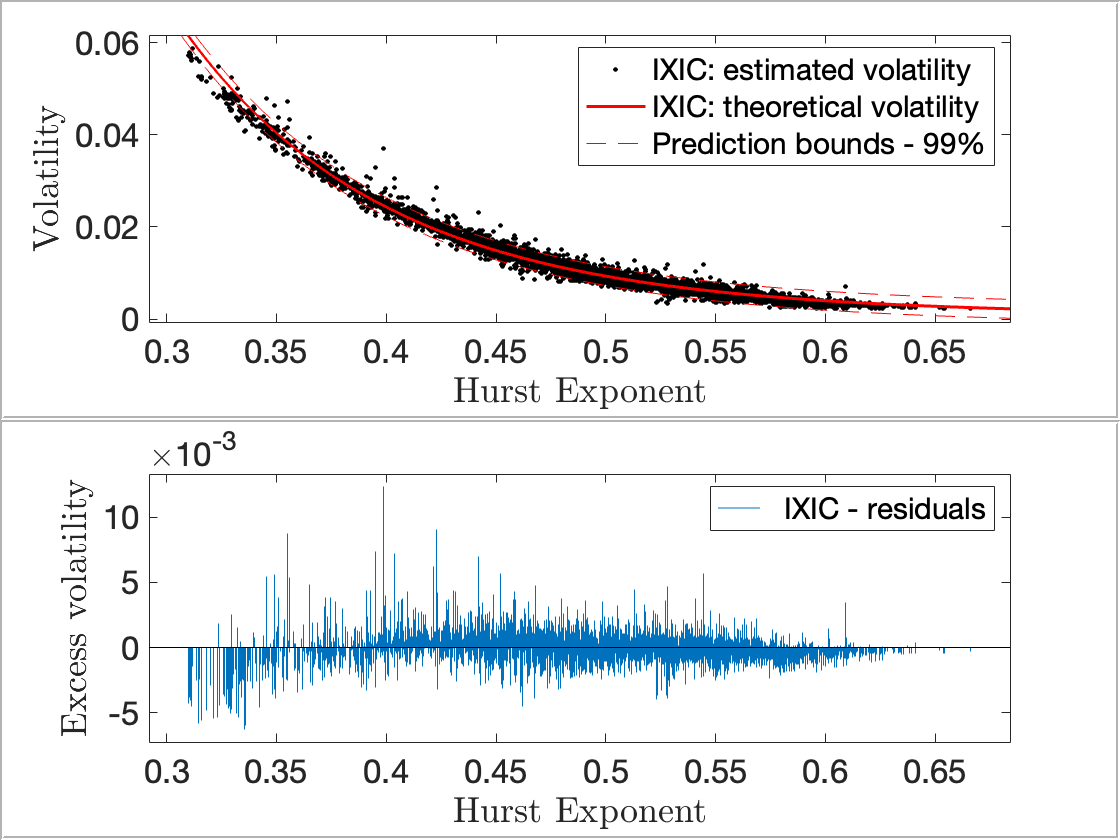}\\
\hspace{-.5cm}\includegraphics[scale=.15, trim=0pt 0pt 100pt 0pt]{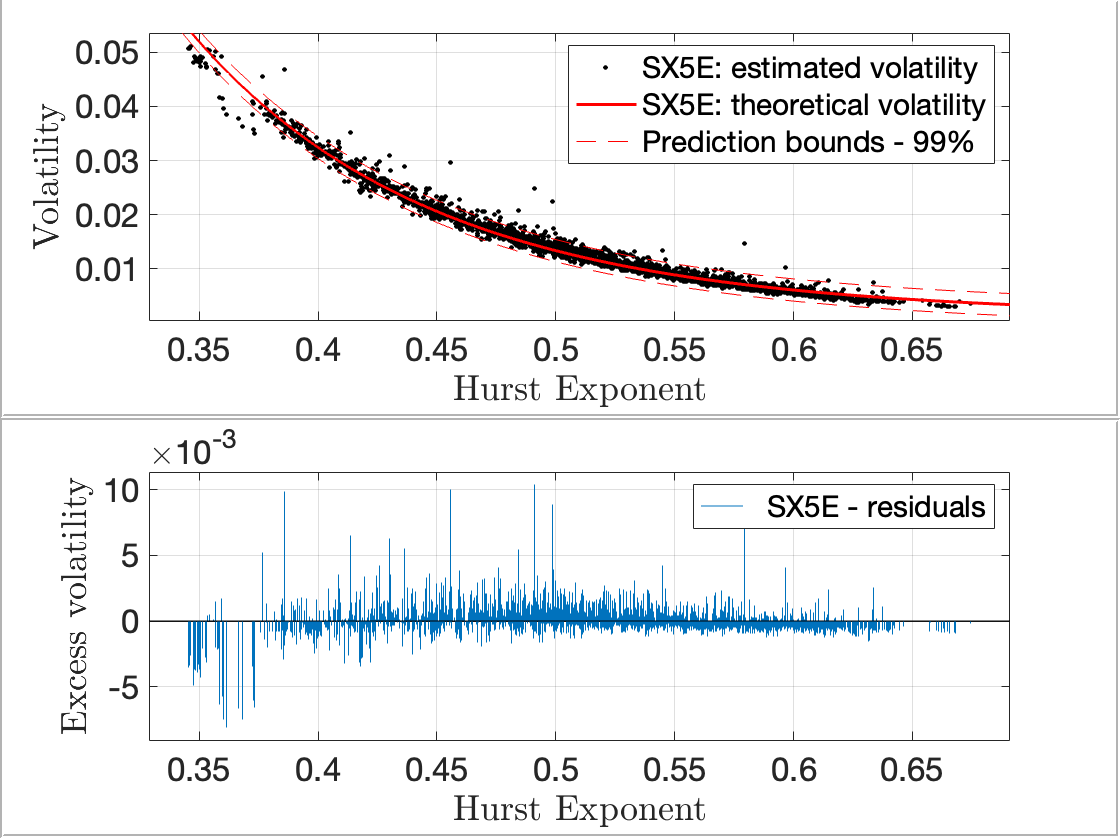}
\includegraphics[scale=.15, trim=0pt 0pt 100pt 0pt]{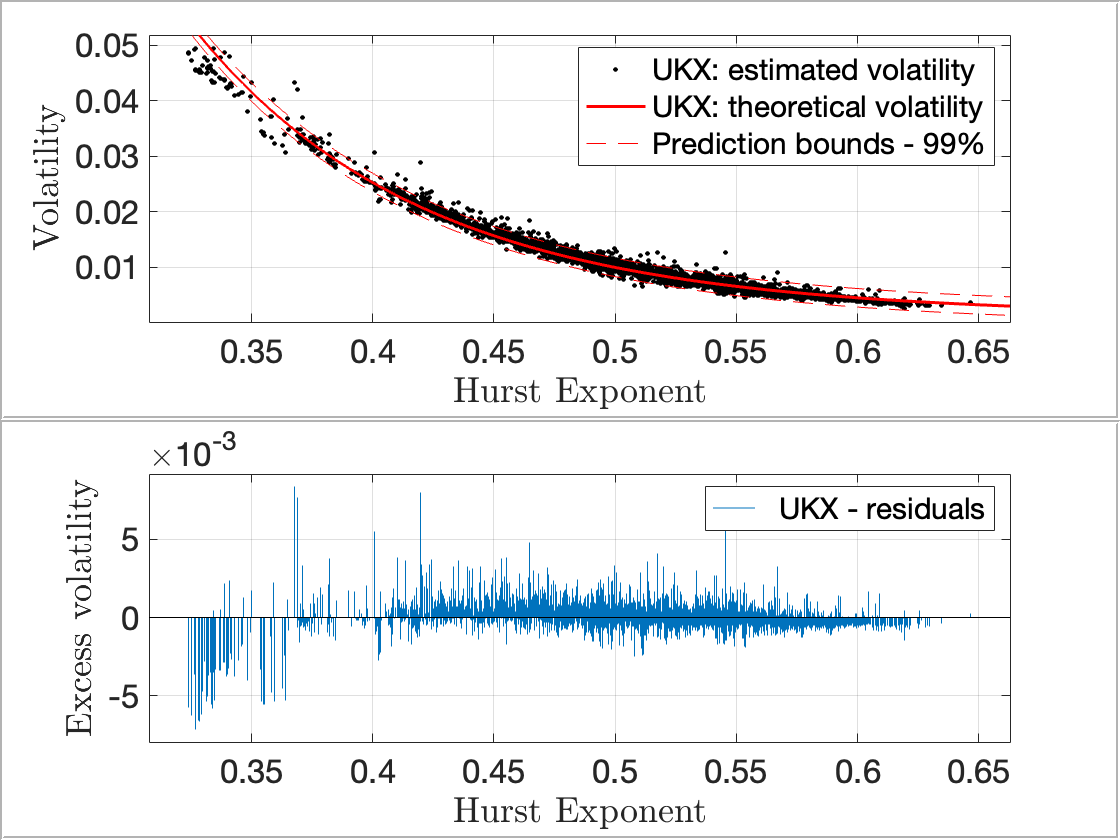}
\includegraphics[scale=.15, trim=0pt 0pt 100pt 0pt]{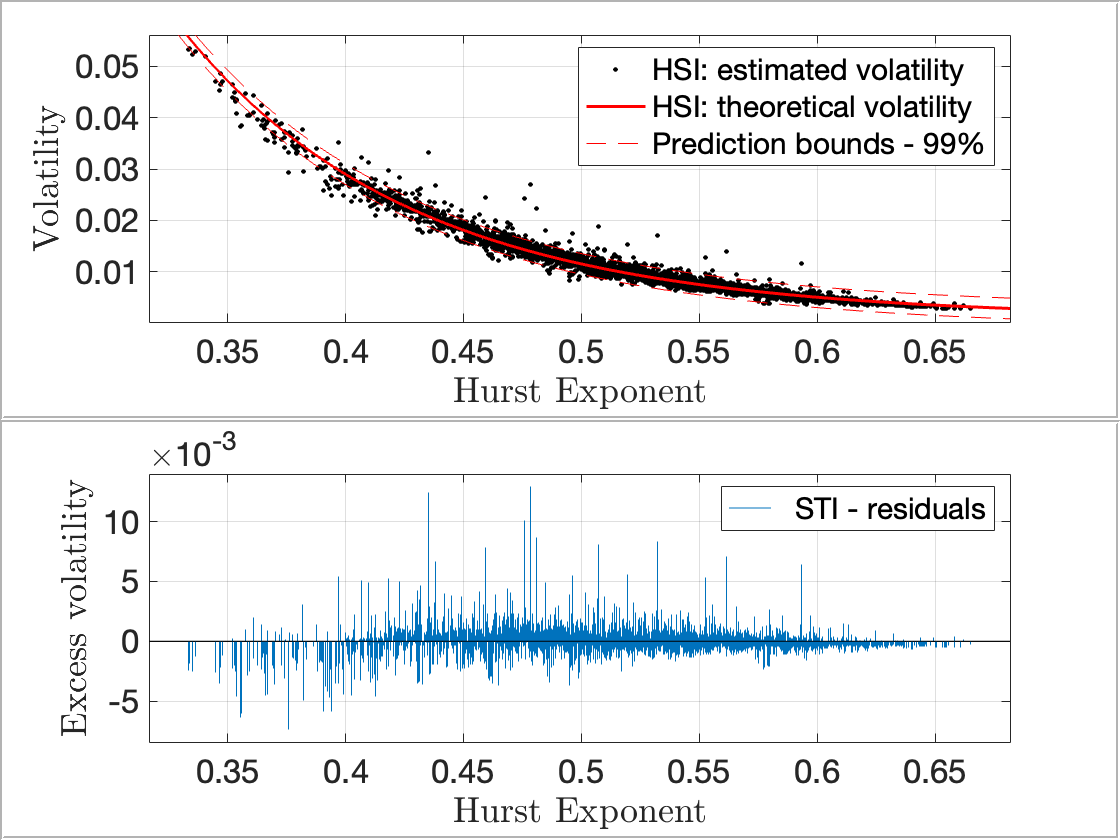}\\
\hspace{-.5cm}\includegraphics[scale=.15, trim=0pt 0pt 100pt 0pt]{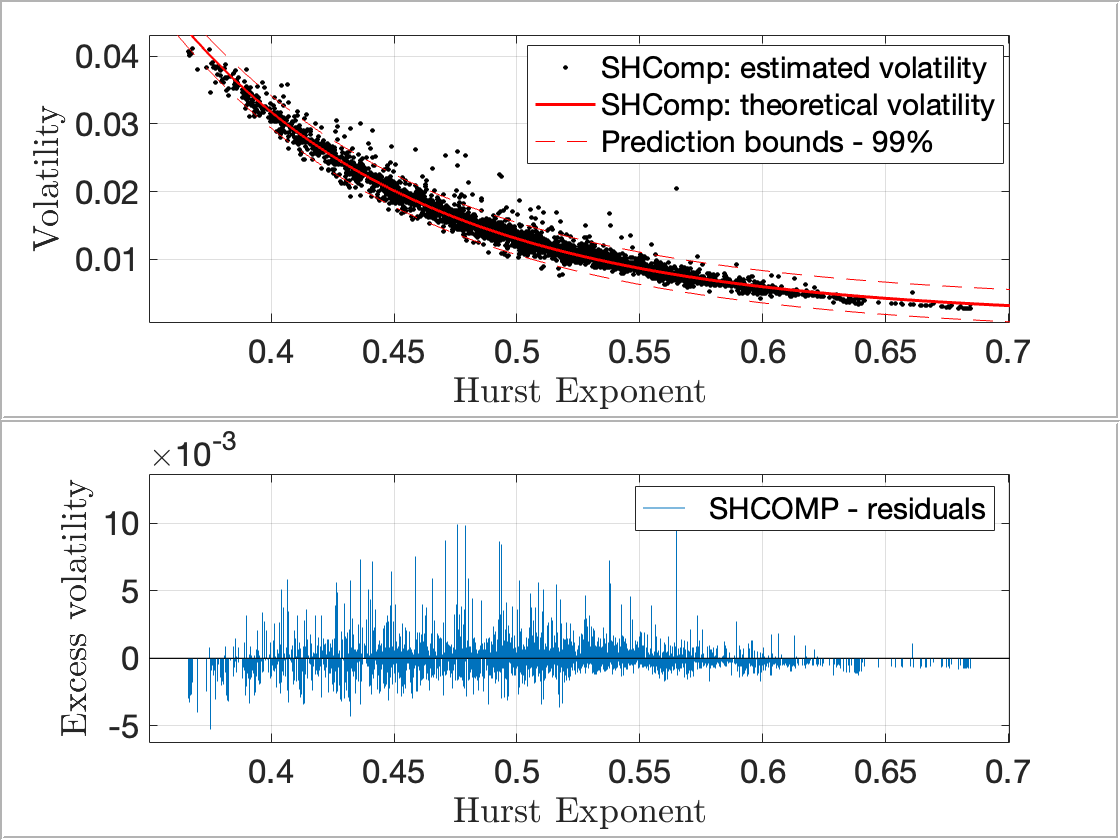}
\includegraphics[scale=.15, trim=0pt 0pt 100pt 0pt]{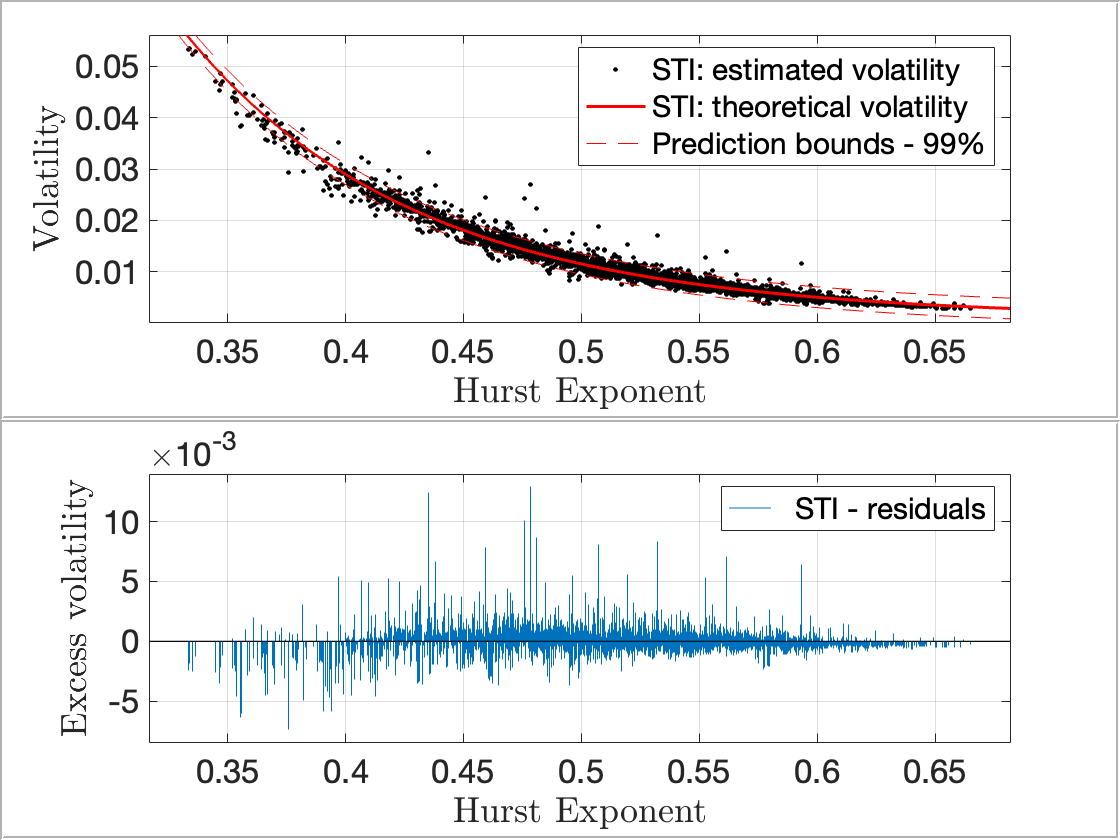}
\includegraphics[scale=.15, trim=0pt 0pt 100pt 0pt]{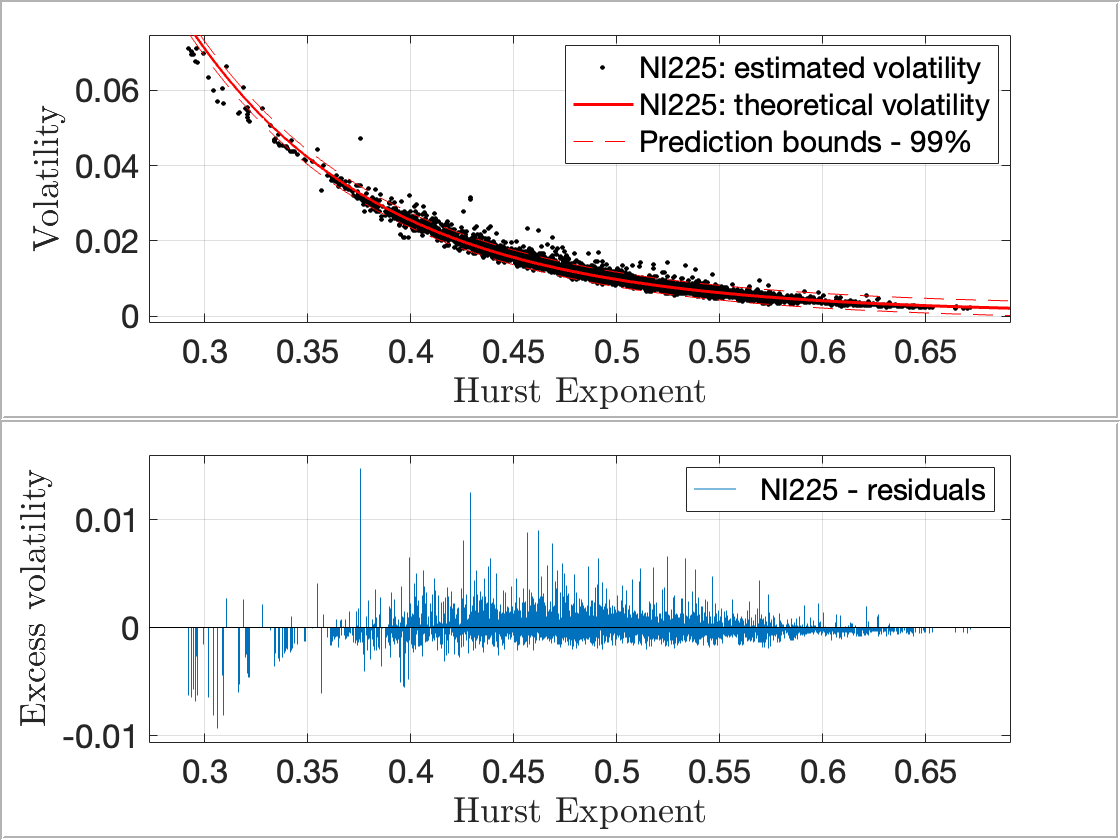}\\
\caption{Estimated Hurst exponents and realized volatility of the one-day log-changes for nine stock indexes and different sampling periods. \textbf{Top panel}: $X$-axis: estimated Hurst exponent $H_t$; $Y$-axis: estimated volatility (black dots), theoretical relation given by equation (\ref{eq:sigmaH2}) (red line), along with the $99\%$ prediction bounds (dotted red lines). Both $H_t$ and realized volatility are estimated with a rolling window of $20$ trading days. The accordance with the theoretical standard deviation expressed by the square root of (\ref{eq:sigmaH2}) seems ubiquitous with respect to time and markets examined. \textbf{Bottom panel}: $X$-axis: estimated Hurst exponent $H_t$; $Y$-axis: residuals of the volatility with respect to the theoretical value. When the Hurst exponent is very low the residuals appear systematically negative for any examined index. This may be due to the fact that both the estimators (of $H_t$ and $\sigma_t$), working on a 20-data points window, do not succeed in estimating the true parameter when data change abruptly. In fact, extremely low values of $H_t$ (resp., extremely high values of $\sigma_t$) typically indicate large and sudden drops in the index which often occur in connection with particularly catastrophic events such as flash crashes or market collapses. In these circumstances, the assumption of a constant $H_t$ (resp. $\sigma_t$), i.e. of local Gaussianity of the log-index process, does not seem reasonable and the impact of sudden and large movements on subsequent estimates decays more slowly.}
\label{fig:Hvolat}
\end{figure}
Table \ref{tab:FitStat} and Figure \ref{fig:Hvolat} summarize the results of the analysis that tests relationship (\ref{eq:sigmaH2}) between the value $H$ and the process standard deviation, both estimated in a neighborhood of $t$. For each rolling window, the estimated volatility $\hat{\sigma}_{t,\delta}$ is plotted versus the estimated Hurst exponent (black dots) and data are fitted using the theoretical relation provided by equation (\ref{eq:sigmaH2}) (red curve) adjusted for the scale. Denoted by $N$ the sample size, the interpolating curve is therefore calculated as  
\begin{equation} \label{eq:Implemented}
 \sigma_{t,\delta} = a + N^{-b\hat{H}_t^{(\delta)}} \left( \frac{\Gamma\left(b\hat{H}_t^{(\delta)}\right)\Gamma\left(1-b\hat{H}_t^{(\delta)}\right)}{\pi \Gamma\left(1+2b\hat{H}_t^{(\delta)}\right)}\right)^{1/2}.
\end{equation}
The fit uses two parameters: a $Y$-axis shifting parameter $a$ and a $X$-axis shifting parameter $b$; a good fit would result in values of $a$ close to $0$ (no vertical shift) and values of $b$ close to $1$ (since parameter $b$ multiplies $\hat{H}_t^{(20)}$, $b=1$ means that no further adjustment is needed for the estimated Hurst exponent to fit relation (\ref{eq:sigmaH2})). For all the indexes, the goodness of fit is quite remarkable, since the values of R squared are all in the range $0.976-0.988$ and the Root Mean Squared Error (RMSE) is in the range $0.0006813-0.0013347$. Also, parameter $a$ is very close to zero (its largest value is 0.001257 for the HSI) and parameter $b$ is always very close to $1$. This indicates that theoretical relation (\ref{eq:sigmaH2}) indeed holds: data are consistent with the conclusion that the examined indexes, within each window, have the same unit-lag variance of an fBm of parameter estimated by $\hat{H}_t^{(20)}$. The fit suggests that the locally asymptotically relation (\ref{eq:LASS1}) holds and that the accordance with the theoretical model is ubiquitous with respect to time and markets.
At extreme and low estimated values of $H_t$, we frequently observe a slight negative bias. This can be attributed to the following explanation: the estimator of $H_t$ presupposes that the data within the window of length $\nu$ should adhere to a normal distribution, since the tangent process is the fBm and it is Gaussian. However, extreme and low values of $H_t$ typically indicate significant disruptions, during which pointwise regularity undergoes significant changes, thereby undermining the assumption of normality. Furthermore, if these values correspond to actual jumps, their influence, combined with the sliding-window filter, tends to result in an underestimation of the subsequent proximate values of $H_t$. \\

Through the empirically validated relationship (\ref{eq:sigmaH2}), it becomes possible to associate each time series with a level of \textit{fair} volatility, corresponding to the benchmark value $H(t)=1/2$. This specific value marks the point at which the process exhibits martingale behavior, thereby aligning with the conditions of market efficiency. As such, it offers a principled approach to addressing a common question in practice: is the observed volatility excessively high or unusually low at a given time? Traditionally, answers to this question have relied on practitioner intuition, portfolio diagnostics, or comparisons with implied volatility levels. However, a more nuanced and theoretically grounded assessment can be made by evaluating the deviation of the observed volatility from the level that would prevail if $H(t)$ was actually $1/2$. This benchmark reflects the equilibrium level toward which market regularity tends, in order to offset transient inefficiencies. Larger volatility levels indicate heightened irregularity (associated with negative autocorrelation), a condition that markets typically correct relatively quickly. In contrast, smaller volatility levels reflect increased regularity (linked to positive autocorrelation), which tends to persist over longer horizons. Insights from behavioral finance -- particularly the cognitive and emotional biases summarized in Table \ref{tab:FinIn} -- help to explain the mechanisms underlying these asymmetric market responses. 
For each index, Table \ref{tab:fair_volatility} reproduces the values of fair volatility along with the 90\%, 95\% and 99\% confidence intervals, deduced by relation \eqref{eq:Implemented}. To ease comparison, also the annualized volatility is calculated using the $T^{1/2}$ rule of thumb. Given that volatility intervals are determined by the choice of the significance level $\alpha$, it would sound more precise to refer to the resulting volatility as the "$\alpha$-fair volatility." \citep{BP2018}


\begin{table}[H]
\centering
\small
\caption{Fair Volatility Estimates and Confidence Intervals (in parentheses annualized volatility and CI)}
{\fontsize{7.54pt}{10pt}\selectfont
\begin{tabular}{lcccc}
\toprule
\textbf{Index} & \textbf{Fair Volatility} & \textbf{90\%-CI} & \textbf{95\%-CI} & \textbf{99\%-CI} \\
\midrule
DJI    & 0.0117 \ (18.6) & (0.0091, 0.0152) \ (14.4, 24.1) & (0.0087, 0.0160) \ (13.8, 25.3) & (0.0079, 0.0177) \ (12.5, 28.0) \\
SPX    & 0.0077 \ (12.2) & (0.0060, 0.0099) \hspace{.09cm} \ (9.5, 15.7) & (0.0058, 0.0104) \hspace{.09cm}\ (9.2, 16.5) & (0.0053, 0.0114) \hspace{.09cm}\ (8.4, 18.0) \\
IXIC   & 0.0093 \ (14.8) & (0.0073, 0.0120) \ (11.5, 19.0) & (0.0069, 0.0126) \ (10.9, 20.0) & (0.0064, 0.0140) \ (10.1, 22.2) \\
SX5E   & 0.0133 \ (21.1) & (0.0104, 0.0172) \ (16.5, 27.3) & (0.0099, 0.0181) \ (15.7, 28.7) & (0.0090, 0.0200) \ (14.2, 31.7) \\
UKX    & 0.0099 \ (15.7) & (0.0078, 0.0128) \ (12.3, 20.3) & (0.0074, 0.0134) \ (11.7, 21.2) & (0.0068, 0.0148) \ (10.7, 23.4) \\
HSI    & 0.0120 \ (19.0) & (0.0095, 0.0153) \ (15.0, 24.2) & (0.0091, 0.0161) \ (14.4, 25.5) & (0.0084, 0.0177) \ (13.3, 28.0) \\
SHCOMP & 0.0130 \ (20.6) & (0.0101, 0.0168) \ (16.0, 26.6) & (0.0097, 0.0177) \ (15.3, 28.0) & (0.0088, 0.0195) \ (13.9, 30.9) \\
STI    & 0.0115 \ (18.3) & (0.0089, 0.0149) \ (14.1, 23.6) & (0.0085, 0.0156) \ (13.4, 24.7) & (0.0078, 0.0172) \ (12.3, 27.3) \\
NI225  & 0.0097 \ (15.4) & (0.0075, 0.0125) \ (11.9, 19.8) & (0.0072, 0.0131) \ (11.4, 20.7) & (0.0066, 0.0145) \ (10.4, 23.0) \\
\bottomrule
\end{tabular}
}
\label{tab:fair_volatility}
\normalsize
\end{table}


\subsection{Empirical key findings}
The empirical analysis yields the following key findings: 
\begin{itemize}[leftmargin=*]
  \item[a)] Markets alternate between periods of efficiency and periods of \textit{positive} (momentum market, $H_t>1/2$) or \textit{negative} (reversal market, $H_t<1/2$) inefficiency (see Figure \ref{fig:Hestim0}). This pattern is consistently present and characteristic across all examined markets.
  \item[b)] The estimates of the pointwise Hurst-H\"older exponent closely align with the theoretical relationship between the Hurst exponent and the variance of a process exhibiting local behavior akin to a fractional Brownian motion (see Figure \ref{fig:Hvolat} and Table \ref{tab:FitStat}). This strongly suggests that stock markets can be effectively modeled through a multifractional process driven by a suitable function $H_t$, which is a reformulation of the statement that log-volatility can be modelled by a fractional Ornstein-Uhlenbeck process (see, e.g., \cite{ComteRenault1998, Gatheral2018} and subsequent strand of research).
  \item[c)] The fact that the estimated sequences $H_t$ are very rough and centered on approximately $1/2$ is highly consistent from a financial perspective, because it suggests that markets tend to recover efficiency when they depart from it for some reason and equilibrium itself is nothing but the collection of disequilibria of opposite signs. In this sense,  for each financial time series, the \textit{fair} volatility can be intended as the value of volatility corresponding to $H=1/2$.
  \item[d)] Given the relationship between $H_t$ and volatility, the apparent roughness of the estimated $H_t$ and its mean-reverting behavior (see Figure \ref{fig:Hestim0}) suggest that the parameter of roughness of the log-volatility could indeed be very low, as found in many recent contributions. However, this result would require a much more in-depth study of how the non linearities appearing in the estimation of $H_t$ for both the log-prices and $H_t$ itself may affect the conclusion of a \textit{rough} regularity-volatility \citep{AngeliniBianchi2023}.

\end{itemize}

\section{Conclusion} \label{sec:Conclusion}
The theoretical arguments and empirical evidence presented in this work indicate that the Hurst-Hölder pointwise exponent constitutes a more informative measure of financial risk. This measure can effectively replace volatility without any loss of informational content, particularly when the process modeling price dynamics exhibits local behavior analogous to fractional Brownian motion.

Adopting the Hurst-Hölder exponent as an alternative to volatility offers several distinct advantages. First and foremost, while volatility alone fails to account for path roughness or smoothness, the Hurst-Hölder exponent explicitly quantifies the degree of pointwise irregularity and relates it to the deviations from the semi-martingale condition, which is characterized by a specific and fixed irregularity pattern reflected in the behavior of the total and quadratic variations. In this sense, the Hurst-Hölder exponent not only quantifies the intensity of randomness affecting a stochastic process but also provides additional insights beyond the unidimensional perspective offered by variability alone.

Second, unlike volatility, which derives meaning primarily from its temporal variations, the Hurst-Hölder exponent is a level-based measure. Its absolute value carries intrinsic significance, making it informative even in static or short-term contexts.

Third, the financial interpretation of the Hurst-Hölder exponent naturally bridges the gap between efficient market theory and behavioral finance. Rather than treating these frameworks as mutually exclusive, the exponent accommodates both perspectives by characterizing them as alternating market phases that dynamically succeed one another within the overall equilibrium seeked by the return of the pointwise Hurst-H\"older exponent to the value $\frac{1}{2}$ characterizing the martingale behavior.

\vspace{.5cm}
\noindent \textbf{Acknowledgments}
The Authors wish to thank the anonymous Reviewers for their valuable comments and suggestions which have helped improving the quality of the work.

\vspace{10pt}
\noindent \textbf{Funding}. This research was supported by Sapienza University of Rome under Grant No. RM120172B346C021.\\

\newpage
\noindent\large{\textbf{Annex 1}}\\
\normalsize
\vspace{.2cm}\\
\vspace{.5cm}\\
%
\textbf{(a) Proof of identity $\frac{\Gamma(1-2H)}{\pi H} \cos (\pi H) = \frac{\Gamma(H)\Gamma(1-H)}{\pi \Gamma(1+2H)}$}.\\
\vspace{.05cm}\\
The proof follows from applying the reflection identity $\Gamma(z)\Gamma(1-z) = \frac{\pi}{\sin(\pi z)}$ with $z=2H$, the double-angle identity for sine and simplifying properly. It follows that $\Gamma(1-2H)=\frac{\pi}{2\sin(\pi H) cos(\pi H) \Gamma(2H)}$, therefore
\begin{eqnarray} \label{eq:identity3}
    \frac{\Gamma(1-2H)}{\pi H} \cos (\pi H) &=& \frac{1}{2H \sin(\pi H)\Gamma(2H)} \\
    &=&\frac{\Gamma(H)\Gamma(1-H)}{\pi \Gamma(1+2H)} \nonumber.
\end{eqnarray}
where the last equality directly follows from solving the Euler's reflection identity with respect to $\sin(\pi H)$.
In the application of Section \ref{sec:Application}, we will test data with respect to $V_H=\frac{\Gamma(H)\Gamma(1-H)}{\pi \Gamma(1+2H)}$.
\begin{flushright}
\qedsymbol{}
\end{flushright}
%
\vspace{.5cm}
\textbf{(b) Proof of identity $\frac{\Gamma(1-2H)}{\pi H} \cos (\pi H) = \frac{\Gamma(2-2H)}{\pi H(1-2H)}\cos(\pi H)$}.
\vspace{.5cm}\\
The identity directly follows from multiplying and dividing by $(1-2H)$ and applying the property $z\Gamma(z)=\Gamma(z+1)$ with $z=1-2H$:
\begin{eqnarray*}
    \frac{\Gamma(1-2H)}{\pi H} \cos (\pi H) &=& \frac{(1-2H)\Gamma(1-2H)}{\pi H (1-2H)} \cos (\pi H)  \\
    &=& \frac{\Gamma(2-2H)}{\pi H (1-2H)} \cos (\pi H)
\end{eqnarray*}
\begin{flushright}
\qedsymbol{}
\end{flushright}
\vspace{0.5cm}
%
\textbf{(c) Proof of identity $\frac{\Gamma(1-2H)}{\pi H} \cos (\pi H)=\frac{1}{2H\sin(\pi H)\Gamma(2H)}=:\frac{A(H)}{\Gamma(H+1/2)^2}$}.\\
\vspace{.05cm}\\
See equation \eqref{eq:identity3}. Here we just observe that by definition $A(H)=\Gamma\left(H+\frac{1}{2}\right)^2V_H$ since
\begin{eqnarray*}
    V_H=\frac{\Gamma(1-2H)}{\pi H} \cos (\pi H) 
    &=& \frac{1}{2H\sin(\pi H)\Gamma(2H)}\\
    &=:&\frac{A(H)}{\Gamma(H+1/2)^2}
\end{eqnarray*}
$A(H)$ is deduced from \eqref{eq:fbm01} and appears in eqs. (28) and (29) of \citep{Lobodaetal2021}, p.31-32. Following \citep{Mishura2008}, it forces the unit-time variance of the process to be equal to 1. Since we are interested in linking the variance of the unit-time increment of the process to the value of $H$ when this changes through time, we will consider the value $V_H$ and will test relation \eqref{eq:VH} where $H$ will be replaced by $H(t)$.
\begin{flushright}
\qedsymbol{}
\end{flushright}

\newpage
\noindent\large{\textbf{Annex 2}}\\
\normalsize
\vspace{.2cm}\\
Let us prove equality 
\begin{equation} \label{eq:start}
    \Gamma\left(H+\frac{1}{2}\right)^2\frac{\Gamma(1-2H)}{\pi H} \cos (\pi H) = \frac{2^{1-4H}\pi \Gamma(2H)}{H\Gamma(H)^2 \sin(\pi H)} 
\end{equation}
where, to ease notation we have dropped the time dependency of $H$, setting $H=H(0)$.
By the Euler's reflection identity
\begin{equation*}
    \Gamma(1-2H) = \frac{\pi}{\Gamma(2H)\sin(2\pi H)}=\frac{\pi}{2\Gamma(2H)\sin(\pi H) \cos(\pi H)}
\end{equation*}
Therefore 
\begin{equation*}
    \frac{\Gamma(1-2H)}{\pi} \cos (\pi H) = \frac{1}{2\Gamma(2H)\sin(\pi H)}
\end{equation*}
Substituting back into \eqref{eq:start}, one has 
\begin{equation} \label{eq:rep}
    \Gamma\left(H+\frac{1}{2}\right)^2\frac{\Gamma(1-2H)}{\pi H} \cos (\pi H)  = \frac{\Gamma\left(H+\frac{1}{2}\right)^2}{2H\Gamma(2H)\sin(\pi H)}
\end{equation}
Exploiting the Legendre's duplication identity
\begin{equation*}
    \Gamma(H)\Gamma\left(H+\frac{1}{2}\right) = 2^{1-2H}\sqrt{\pi}\Gamma(2H)
\end{equation*}
it follows
\begin{equation*}
    \Gamma\left(H+\frac{1}{2}\right)^2 = \frac{2^{2(1-2H)}\pi \Gamma(2H)^2}{\Gamma(H)^2}
\end{equation*}
i.e., replacing into the numerator of \eqref{eq:rep}, it follows
\begin{equation*}
    \Gamma\left(H+\frac{1}{2}\right)^2\frac{\Gamma(1-2H)}{\pi H} \cos (\pi H) = \frac{2^{1-4H}\pi\Gamma(2H)}{H\Gamma(H)^2\sin(\pi H)}.
\end{equation*}
\begin{flushright}
\qedsymbol{}
\end{flushright}

\newpage
\noindent\large{\textbf{Annex 3}}\\
\normalsize
\vspace{.2cm}\\
Let $H \sim \mathcal{N}\left(\frac{1}{2},\sigma^2_H\right)$, with $E(H)=\frac{2^{1-4H}\pi\Gamma(2H)}{H\Gamma(H)^2\sin(\pi H)}$. The goal is to find the expected value $\mathbb{E}[E(H)]$ and since the variance is small, it is reasonable to think about a Taylor expansion around the mean value.\\
\begin{equation*}
    E(H)=E\left(\frac{1}{2}\right)+ E'\left(\frac{1}{2}\right)\left(H-\frac{1}{2}\right)+\frac{1}{2}E''\left(\frac{1}{2}\right)\left(H-\frac{1}{2}\right)^2+\mathcal{O}\left(\left(H-\frac{1}{2}\right)^3\right).
\end{equation*}
Therefore,
\begin{eqnarray}
\mathbb{E}\left(E(H)\right)&\!\!\!=\!\!\!&E\left(\frac{1}{2}\right)\!+\! E'\left(\frac{1}{2}\right)\!\mathbb{E}\!\left[\left(H\!-\!\frac{1}{2}\right)\right]\!+\!\frac{1}{2}E''\left(\frac{1}{2}\right)\!\mathbb{E}\!\left[\left(H\!-\!\frac{1}{2}\right)^2\right]\!+\!\mathcal{O}\left(\left(H\!-\!\frac{1}{2}\right)^3\right) \nonumber \\
    &\!\!\!=\!\!\!& E\left(\frac{1}{2}\right)+\frac{\sigma^2_H}{2}E''\left(\frac{1}{2}\right) \nonumber \\
    &\!\!\!=\!\!\!& 1 + \frac{\sigma^2_H}{2}E''\left(\frac{1}{2}\right) \nonumber
\end{eqnarray}
\\
First notice that
\begin{equation*}
    \log E(H) = \log \pi +\log 2^{1-4H}+\log \Gamma(2H)-\log H -\log \Gamma(H)^2 - \log\left(\sin(\pi H)\right).
\end{equation*}
Differentiating with respect to $H$:
\begin{equation*}
    \frac{E'(H)}{E(H)}=-4\log2 +2\psi(2H)-\frac{1}{H}-2\psi(H)-\pi\cot(\pi H)
\end{equation*}
where $\psi(H)$ is the digamma function defined as $\frac{\Gamma'(H)}{\Gamma(H)}$.\\
Therefore 
\begin{eqnarray}
    E'\left(\frac{1}{2}\right)&=&-4\log2 +2\psi(1)-2-2\psi\left(\frac{1}{2}\right)-\pi\cot\left(\frac{\pi}{2}\right)\nonumber \\
    &=& -4\log2+2(-\gamma)-2-2(-\gamma-2\log2) \nonumber \\
    &=& -2, \nonumber
\end{eqnarray}
where $\gamma$ is the Euler-Mascheroni constant ($\psi(1)=-\gamma$ and $\psi\left(\frac{1}{2}\right)=-\gamma-2\log2$).\\
\vspace{.2cm}\\
\\
To find $E''(H)$ we need to differentiate the logarithmic derivative, i.e.
\begin{equation*}
    E'(H)=E(H)\left[-4\log2 +2\psi(2H)-\frac{1}{H}-2\psi(H)-\pi\cot(\pi H)\right].
\end{equation*}
Therefore,
\begin{eqnarray}
    E''(H) &=& E'(H)\left[-4\log2 +2\psi(2H)-\frac{1}{H}-2\psi(H)-\pi\cot(\pi H)\right] \nonumber \\
    && + E(H)\left[4\psi'(2H)+\frac{1}{H^2}-2\psi'(H)+\pi^2\csc^2(\pi H)\right]
\end{eqnarray}
Since we already know that $E\left(\frac{1}{2}\right)=1$ and $E'\left(\frac{1}{2}\right)=-2$, it follows
\begin{eqnarray}
    E''\left(\frac{1}{2}\right) &=& -2\left[-4\log2 +2\psi(1)-2-2\psi\left(\frac{1}{2}\right)-\pi\cot\left(\frac{\pi}{2}\right)\right] \nonumber \\
    && + \left[4\psi'(1)+4-2\psi'\left(\frac{1}{2}\right)+\pi^2\csc^2\left(\frac{\pi}{2}\right)\right] \nonumber
\end{eqnarray}
Since:
\begin{itemize}
    \item the term in the first square brackets above is precisely $E'\left(\frac{1}{2}\right)=-2$,
    \item $\psi'(1)=\frac{\pi^2}{6}$ and $\psi'\left(\frac{1}{2}\right)=\frac{\pi^2}{2}$ and $\csc\left(\frac{\pi}{2}\right)=1$
\end{itemize} 
it follows
\begin{equation*}
    E''\left(\frac{1}{2}\right)=8+\frac{2}{3}\pi^2.
\end{equation*}
Using the Taylor expansion approximation, one finally has
\begin{equation*}
    \mathbb{E}(E(H)) \approx 1+\left(4+\frac{\pi^2}{3}\right)\sigma^2_H
\end{equation*}
\begin{flushright}
\qedsymbol{}
\end{flushright}

\newpage
\bibliographystyle{plain} 
\bibliography{Bibliography10}

\end{document}